\documentclass[aps, prd, floatfix, nofootinbib, superscriptaddress, twocolumn]{revtex4-1}

\usepackage{amsmath}
\usepackage{amssymb}
\usepackage{amsfonts}
\usepackage{color}
\usepackage{CJKutf8}
\usepackage{latexsym}
\usepackage{textcomp}

\usepackage[mathscr,scaled=1.15]{urwchancal}
\DeclareFontFamily{OT1}{pzc}{}
\DeclareFontShape{OT1}{pzc}{m}{it}%
{<-> s * [1.15] pzcmi7t}{}
\DeclareMathAlphabet{\mathpzc}{OT1}{pzc}{m}{it}

\usepackage{supertabular}
\usepackage{placeins}
\usepackage{epsfig}
\usepackage{graphicx}
\usepackage{multirow}

\definecolor{purple}{rgb}{0.5,0,0.5}
\definecolor{blue}{rgb}{0.0,0,0.9}
\definecolor{prdblue}{rgb}{0.133,0.118,0.498}
\usepackage[colorlinks=true, pdfstartview=FitV, linkcolor=prdblue, citecolor= prdblue, urlcolor=prdblue]{hyperref}

\newcounter{Afigure}
\newcounter{Atable}

\newenvironment{AAfigure}{\refstepcounter{Afigure}}{}
\newenvironment{AAtable}[1][]{\refstepcounter{Atable}}{}

\begin{document}

\begin{CJK}{UTF8}{song}

\title{$\,$\\[-6.5ex]\hspace*{\fill}{\normalsize{\sf\emph{Preprint nos}.\ NJU-INP 057/22, USTC-ICTS/PCFT-22-11}}\\[1.25ex]
Composition of low-lying $\mathbf{J=\tfrac{3}{2}^\pm \,\Delta}$-baryons}

\date{2022 March 22}

\author{Langtian Liu} 
\affiliation{School of Physics, Nanjing University, Nanjing, Jiangsu 210093, China}
\affiliation{Institute for Nonperturbative Physics, Nanjing University, Nanjing, Jiangsu 210093, China}
\author{Chen Chen} 
\email[]{chenchen1031@ustc.edu.cn}
\affiliation{Interdisciplinary Center for Theoretical Study, University of Science and Technology of China, Hefei, Anhui 230026, China}
\affiliation{Peng Huanwu Center for Fundamental Theory, Hefei, Anhui 230026, China}
%
\author{Ya Lu} 
\affiliation{School of Physics, Nanjing University, Nanjing, Jiangsu 210093, China}
\affiliation{Institute for Nonperturbative Physics, Nanjing University, Nanjing, Jiangsu 210093, China}
\affiliation{Department of Physics, Nanjing Tech University, Nanjing 211816, China}
\author{Craig D.~Roberts}
\email[]{cdroberts@nju.edu.cn}
\affiliation{School of Physics, Nanjing University, Nanjing, Jiangsu 210093, China}
\affiliation{Institute for Nonperturbative Physics, Nanjing University, Nanjing, Jiangsu 210093, China}
\author{Jorge Segovia}
\affiliation{Dpto.\ Sistemas F\'isicos, Qu\'imicos y Naturales, Univ.\ Pablo de Olavide, E-41013 Sevilla, Spain}
\affiliation{Institute for Nonperturbative Physics, Nanjing University, Nanjing, Jiangsu 210093, China}

\begin{abstract}
A Poincar\'e-covariant quark+diquark Faddeev equation is used to develop insights into the structure of the four lightest $(I,J^P=\tfrac{3}{2},\tfrac{3}{2}^\pm)$ baryon multiplets.  Whilst these systems can contain isovector-axialvector and isovector-vector diquarks, one may neglect the latter and still arrive at a reliable description.
The $(\tfrac{3}{2},\tfrac{3}{2}^+)$ states are the simpler systems, with features that bear some resemblance to quark model pictures, \emph{e.g}., their most prominent rest-frame orbital angular momentum component is $\mathsf S$-wave and the $\Delta(1600)\tfrac{3}{2}^+$ may reasonably be viewed as a radial excitation of the $\Delta(1232)\tfrac{3}{2}^+$.
The $(\tfrac{3}{2},\tfrac{3}{2}^-)$ states are more complex: the $\Delta(1940)\tfrac{3}{2}^-$ expresses little of the character of a radial excitation of the $\Delta(1700)\tfrac{3}{2}^-$; and whilst the rest-frame wave function of the latter is predominantly $\mathsf P$-wave, the leading piece in the $\Delta(1940)\tfrac{3}{2}^-$ wave function is $\mathsf S$-wave, in conflict with quark model expectations.
Experiments that can test these predictions, such as large momentum transfer resonance electroexcitation, may shed light on the nature of emergent hadron mass.
\end{abstract}

\maketitle

\end{CJK}


%
\section{Introduction}
Questions relating to the composition of baryons have been asked for roughly one hundred years.  Answers possessing an appealing simplicity within the framework of quantum mechanics were provided by the (constituent) quark model \cite{Gell-Mann:2015noa} via its progeny, \emph{viz}.\ three-body potential models \cite{Capstick:2000qj, Crede:2013kia, Giannini:2015zia, Plessas:2015mpa, Eichmann:2022zxn}.  In such models, baryons constituted from combinations of up ($u$), down ($d$), and strange ($s$) valence quark flavours can be grouped into multiplets of SU$(6)\otimes$O$(3)$,
labelled by their flavour content, spin, and orbital angular momentum.  From this perspective, the four lightest $(I,J^P=\tfrac{3}{2},\tfrac{3}{2}^\pm)$ $\Delta$-baryons, built from isospin $I=\tfrac{3}{2}$ combinations of three $u$ and/or $d$ quarks, are typically viewed as follows:
$\Delta(1232)\tfrac{3}{2}^+$, ${\mathsf S}$-wave ground-state;
$\Delta(1600)\tfrac{3}{2}^+$, radial excitation of the $\Delta(1232)$, hence, ${\mathsf S}$-wave;
$\Delta(1700)\tfrac{3}{2}^-$, $L=1$ orbital angular momentum excitation of the $\Delta(1232)$, so, ${\mathsf P}$-wave;
and $\Delta(1940)\tfrac{3}{2}^-$, radial excitation of the $\Delta(1700)$, thus, also ${\mathsf P}$-wave.

Quark models are practically useful in many applications; yet, so far as spectra are concerned, they typically produce masses for radial excitations of the ground-state that are too large when compared with the lowest-mass orbital angular momentum excitation \cite[Sec.\,15]{Zyla:2020zbs}.  The best known example is the Roper resonance, $N(1440)\tfrac{1}{2}^+$, discussed elsewhere \cite{Burkert:2019bhp}, which is predicted to lie above the nucleon's parity partner, $N(1535)\tfrac{1}{2}^-$, in contradiction of experiment.  The same issue is encountered in decuplet baryons with, \emph{e.g}., the calculated mass of the $\Delta(1600)\tfrac{3}{2}^+$ being greater than that of the $\Delta(1700)\tfrac{3}{2}^-$.

Potential models are also challenged by quantum chromodynamics (QCD), which requires a Poincar\'e covariant description of baryon structure that leads to a Poincar\'e invariant explanation of their properties \cite{Brodsky:2022fqy}.
For instance, the evaluation of hadron distribution functions (DFs) requires Poincar\'e covariance in order to ensure, \emph{inter alia}, the proper domain of DF support \cite{Holt:2010vj}; and modern electroproduction experiments are probing ground- and excited-state baryons using photons with virtuality approaching $10 \,m_p^2$ \cite{Carman:2020qmb, Brodsky:2020vco, Mokeev:2022xfo}, where $m_p$ is the proton mass.
Furthermore, whilst the total angular momentum of a bound-state is Poincar\'e-invariant, this is not true of any separation into spin and orbital angular momentum components carried by the system's identified constituents \cite{Coester:1992cg}.  Hence, potential model wave functions might only provide a rudimentary guide to baryon structure.

An alternative lies in calculations of the bound-state pole position and residue in the six-point Schwinger function that describes three-quark--to--three-quark scattering.  This is the matrix element upon which simulations of lattice-QCD focus in order to extract baryon masses \cite{Edwards:2011jj, Fodor:2012gf}.  It is also the basis for studies of baryon composition using continuum Schwinger function methods (CSMs) \cite{Eichmann:2016yit, Burkert:2019bhp, Qin:2020rad}.  Within this framework, the problem is expressed in a Poincar\'e-covariant three-body Faddeev equation whose solution yields the masses-squared and bound state amplitudes of all baryons in the channel under consideration.  Baryon spectra and dynamical properties have been computed \cite{Eichmann:2009qa, Eichmann:2011vu, Wang:2018kto, Qin:2019hgk} at leading-order (rainbow-ladder) in a systematic, symmetry-preserving truncation scheme
\cite{Qin:2013mta, Qin:2014vya, Binosi:2016rxz}; and efforts are underway to implement more sophisticated truncations \cite{Qin:2020jig}.

\begin{figure}[t]
\centerline{%
\includegraphics[clip, width=0.48\textwidth]{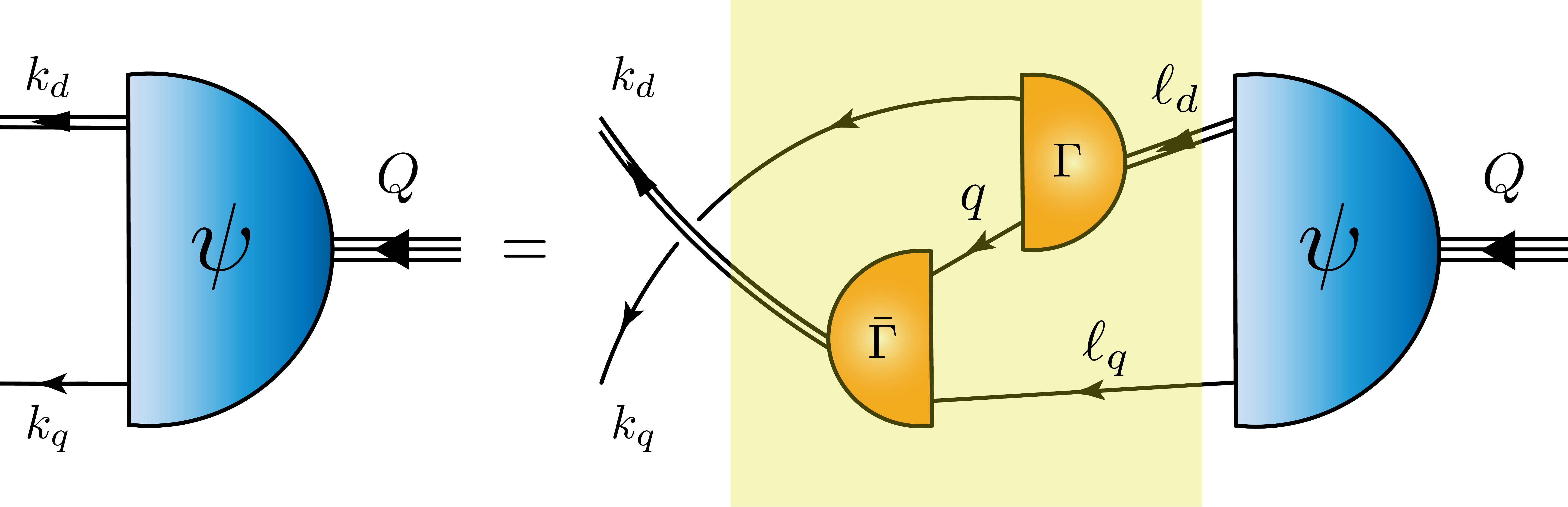}}
\caption{\label{FigFaddeev}
Quark+diquark Faddeev equation, a linear integral equation for the Poincar\'e-covariant matrix-valued function $\psi$, the Faddeev amplitude for a baryon with total momentum $Q=\ell_q+\ell_d=k_q+k_d$. $\psi$ describes the relative momentum correlation between the dressed-quarks and -diquarks. Legend. \emph{Shaded rectangle} -- Faddeev kernel; \emph{single line} -- dressed-quark propagator, $S(\ell)$; $\Gamma^{J^P}(k;K)$ -- diquark correlation amplitude; and \emph{double line} -- diquark propagator, $D^{J^P}(K)$.
}
\end{figure}

Meanwhile, a simplification of the full three-body problem continues to be employed with success.  Namely, the interacting quark+diquark picture, illustrated in Fig.\,\ref{FigFaddeev}, that was derived from the three-body equation in Refs.\,\cite{Cahill:1988dx, Burden:1988dt, Reinhardt:1989rw, Efimov:1990uz}.  The approximation is efficacious because any interaction that is able to generate Nambu-Goldstone modes as dressed-quark+antiquark bound-states and reproduce the measured value of their leptonic decay constants, must also produce strong colour-antitriplet correlations between any two dressed quarks contained within a hadron \cite{Barabanov:2020jvn}.
In general, for light-quark systems, the following diquark correlations are possible: isoscalar-scalar, $(I,J^P=0,0^+)$; isovector-axialvector; isoscalar-pseudoscalar; isoscalar-vector; and isovector-vector.  Within a given system, channel dynamics determines the relative strengths of these correlations.  Herein, owing to the fact that $I=\tfrac{3}{2}$ baryons cannot be built from $I=0$ diquarks, we just need to consider $(1,1^\pm)$ correlations.

It is worth stressing that the diquark correlations discussed herein are fully dynamical, appearing in a Faddeev kernel which requires their continual breakup and reformation.  Hence, they are very different from the pointlike, static diquarks introduced more than fifty years ago \cite{Anselmino:1992vg} with a view to solving the so-called ``missing resonance'' problem \cite{Aznauryan:2011ub}.  This essentially active character of the valence quarks within diquarks entails that the spectrum produced by Fig.\,\ref{FigFaddeev} possesses a richness that cannot be explained by two-body models, something also found in numerical simulations of lattice-regularised QCD \cite{Edwards:2011jj}.

\begin{figure}[t]

\hspace*{-4ex}
\includegraphics[width=0.39\textwidth]{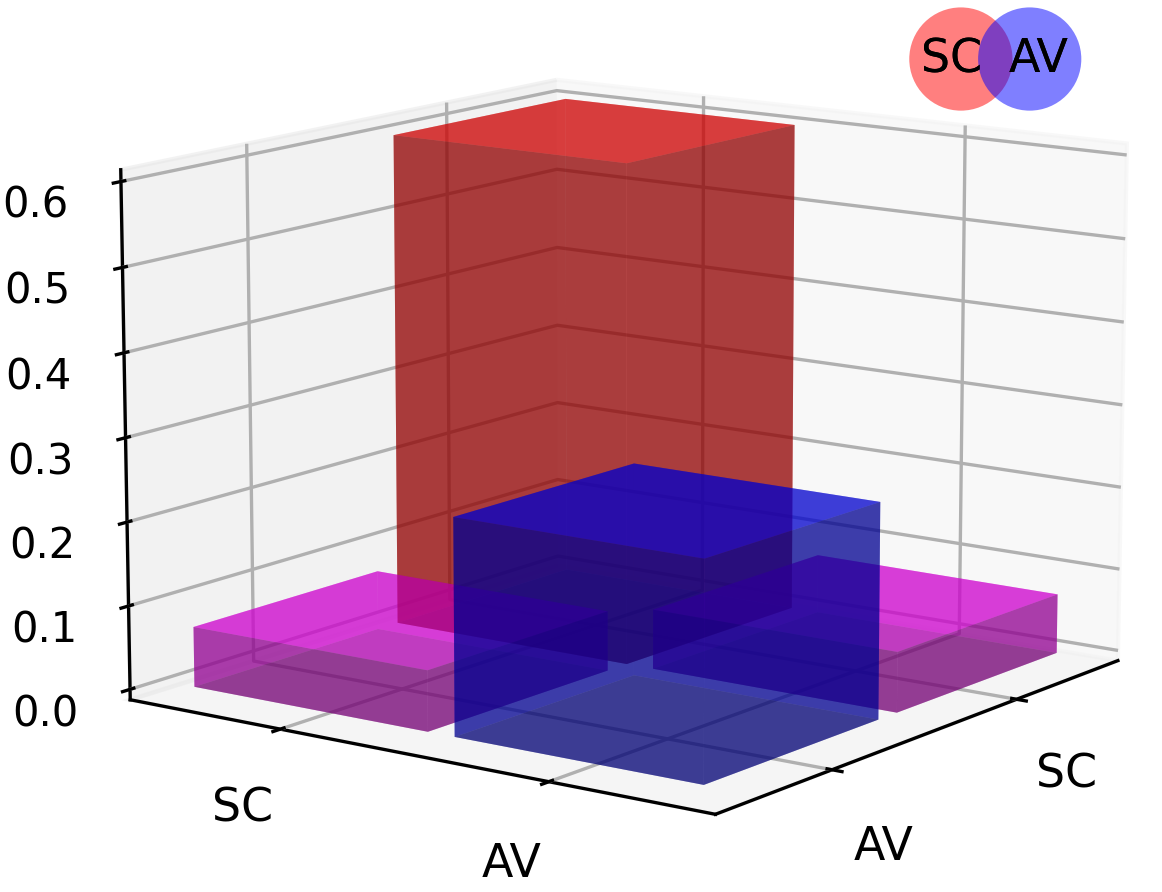}
\caption{\label{FigNucleonqq}
Contributions of the various diquark components to the canonical normalisation of the Poincar\'e-covariant nucleon Faddeev wave function.  Whilst the $[ud]_{0^-}$ isoscalar-scalar diquark (SC) is dominant, material contributions also owe to the $\{uu\}_{1^+}$, $\{ud\}_{1^+}$ isovector-axialvector correlations (AV).
}
\end{figure}

An analysis of the four lowest-lying $(\tfrac{1}{2},\tfrac{1}{2}^\pm)$ baryons -- the nucleon and some kindred systems -- made using the quark+diquark framework is presented elsewhere \cite{Chen:2017pse}.
It was found therein that $(0,0^+)$ and $(1,1^+)$ diquarks dominate the wave functions of the lightest $(\tfrac{1}{2},\tfrac{1}{2}^+)$ doublets.  This is illustrated for the nucleon ground-state in Fig.\,\ref{FigNucleonqq}: roughly 60\% of the proton's canonical normalisation constant is provided by the $(0,0^+)$ correlation, but the remainder owes to the $(1,1^+)$ correlation and constructive $(0,0^+) \otimes (1,1^+)$ interference.
(The canonical normalisation constant is related to the $Q^2=0$ value of the charge form factors associated with the electrically charged members of a given hadron multiplet: in this case, that is the proton Dirac form factor.)
As explained elsewhere \cite{Chang:2022jri, Lu:2022cjx}, the size of the $(1,1^+)$-linked contributions is sufficient to explain the measured ratio of proton valence-quark distribution functions \cite{Abrams:2021xum, Cui:2021gzg}.

\begin{figure}[t]
\vspace*{0ex}

\hspace*{-4ex}\includegraphics[width=0.45\textwidth]{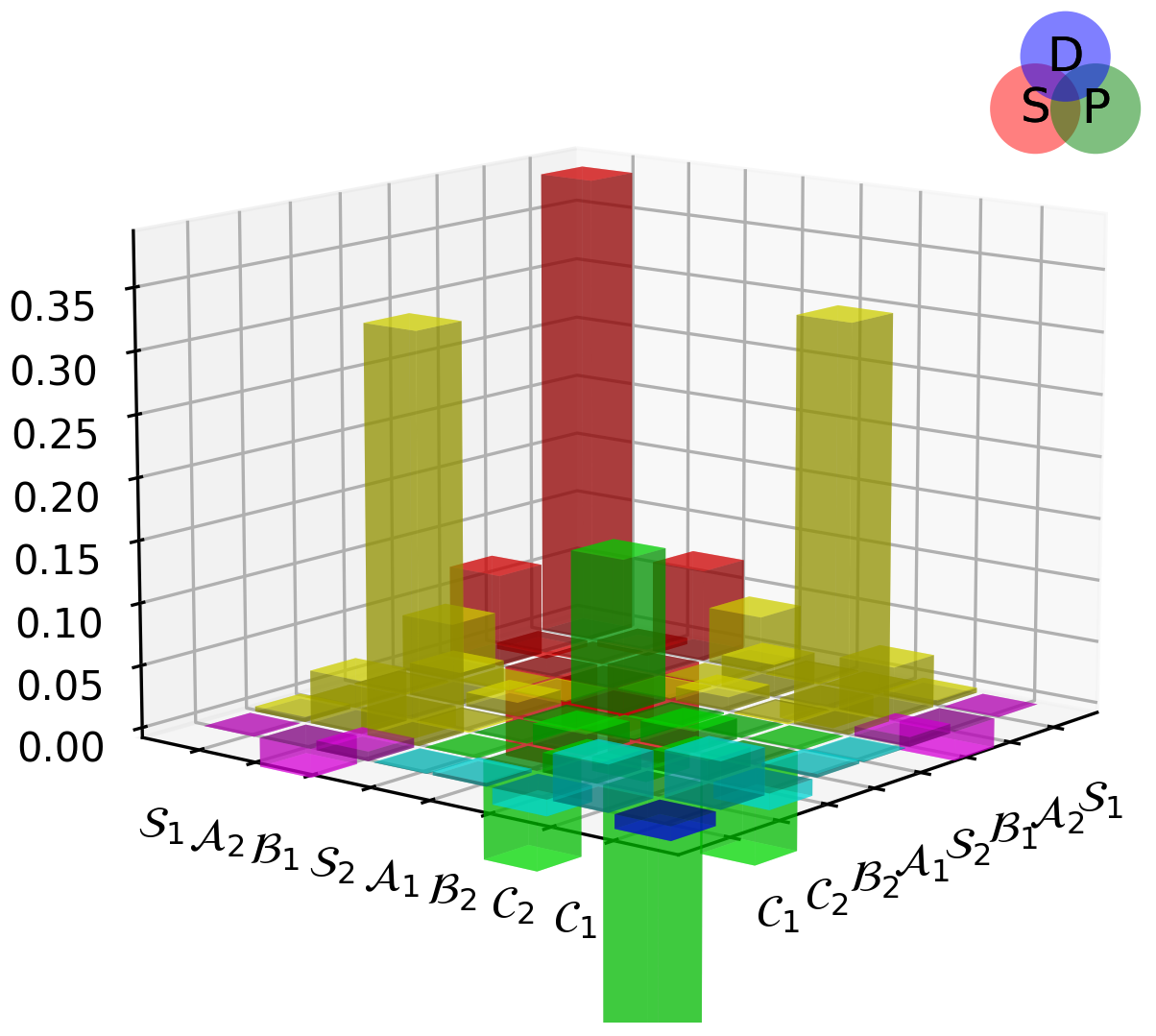}

\vspace*{-51ex}

\leftline{\hspace*{0.5em}{\large{\textsf{A}}}}

\vspace*{46ex}

\leftline{\hspace*{0.5em}{\large{\textsf{B}}}}

\includegraphics[width=0.4\textwidth]{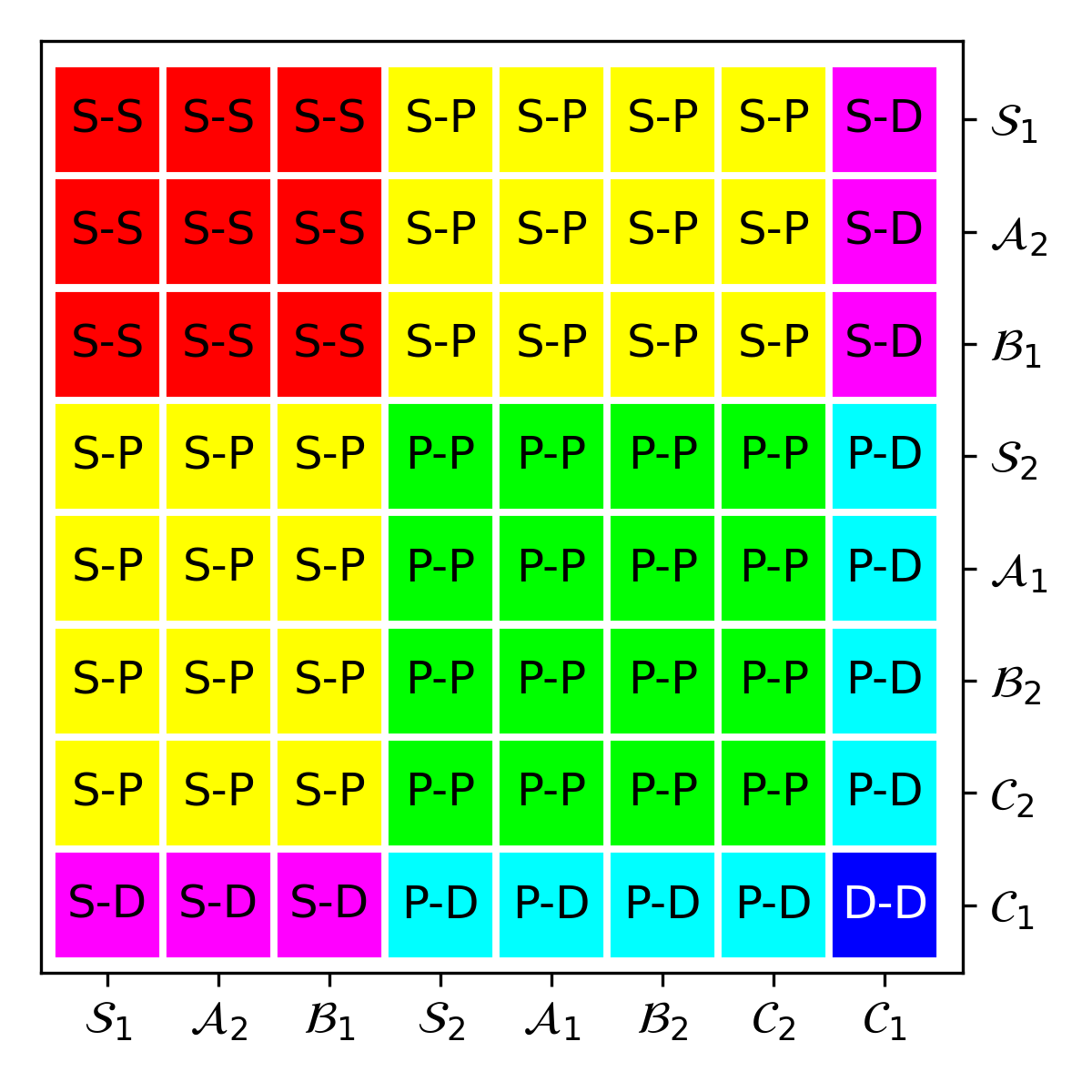}

\caption{\label{FigNucleonL}
\emph{Upper panel}\,--\,{\sf A}.
Contributions of the various quark+diquark orbital angular momentum components to the canonical normalisation of the Poincar\'e-covariant nucleon wave function after projection into the rest frame: there are both positive (above plane) and negative (below plane) contributions to the overall positive normalisation.  The values drawn here are listed in 
Table~A.\hyperref[NewTableSPDN940]{1}
\emph{Lower panel}\,--\,{\sf B}. Legend for interpretation of upper panel, identifying interference between the distinct orbital angular momentum basis components.
Details of the decomposition are provided in Table~\ref{TabL}.  It follows the scheme described in Refs.\,\cite{Oettel:1998bk, Cloet:2007pi} and uses a pictorial representation based on that in Ref.\,\cite{Hilger:2015ora}.
}
\end{figure}

Furthermore, as shown for the nucleon in Fig.\,\ref{FigNucleonL}, projected into the rest frame, these wave functions have significant ${\mathsf S}$-wave components; yet they also contain material ${\mathsf P}$-wave structures and the canonical normalisation receives measurable ${\mathsf S}\otimes {\mathsf P}$-wave interference contributions.
In addition \cite{Chen:2017pse}, the first $\tfrac{1}{2}^+$ excited state may fairly be described as the radial excitation of the ground state.
In these outcomes, there are some parallels with quark model expectations for these states.

On the other hand, the related $(\tfrac{1}{2},\tfrac{1}{2}^-)$ doublets fit a different picture \cite{Chen:2017pse}: $(1,1^-)$ diquarks play an important role; the wave functions are predominantly ${\mathsf P}$-wave in nature, but contain significant ${\mathsf S}$-wave components; and the heavier states are not simply radial excitations of their lighter partners.

Notably, in quantum field theory, all differences between positive- and negative-parity states can be attributed to chiral symmetry breaking, as highlighted by the $\rho$-$a_1$ meson complex \cite{Weinberg:1967kj, Chang:2011ei, Williams:2015cvx, Qin:2020jig}.  In the light-quark sector, such symmetry breaking is almost entirely dynamical.  Dynamical chiral symmetry breaking (DCSB) is a corollary of emergent hadron mass (EHM) \cite{Roberts:2020hiw, Roberts:2021nhw, Aguilar:2021uwa, Binosi:2022djx}; hence, quite probably linked tightly with confinement, which requires a $\sim1$\,fm$^{-1}$ mass-scale to be effective \cite{Horn:2016rip}.  Consequently, experiments that can test predictions made for differences between parity partners in the hadron spectrum are valuable.  These features imbue quantum field theory studies of $(\tfrac{3}{2},\tfrac{3}{2}^\pm)$ baryons with particular interest.

Our approach to the $(\tfrac{3}{2},\tfrac{3}{2}^\pm)$-baryon bound-state problems is sketched in Sec.\,\ref{secFaddeev}.   Solutions for the masses and Poincar\'e-covariant wave functions of the lowest-lying such states are described and dissected in Sec.\,\ref{Solutions}.  Section~\ref{epilogue} provides a summary and perspective.

\section{Bound State Equations}
\label{secFaddeev}
In studying $(\tfrac{3}{2},\tfrac{3}{2}^\pm)$ baryons, we follow the analysis of $(\tfrac{1}{2},\tfrac{1}{2}^\pm)$ states in Ref.\,\cite{Chen:2017pse}.
For instance:
we assume isospin symmetry throughout;
the diquark correlation amplitudes, $\Gamma^{J^P}$, are similar;
the light-quark and diquark propagators, $S$, $D^{J^P}$, are unchanged -- see Appendix\,\ref{APropagators};
and the effective masses of the relevant diquark correlations are (in GeV)
\begin{equation}
\label{diquarkmasses}
m_{\{uu\}_{1^+}}=0.9\,,\;
m_{\{uu\}_{1^-}}=1.4\,.
\end{equation}
The mass splitting here is commensurate with that in the $\rho$-$a_1$ complex \cite{Zyla:2020zbs}.
On the other hand, since the negative-parity diquarks are heavy, we emulate Ref.\,\cite[Sec.\,4.1.4]{Yin:2021uom} in electing not to include the $g_{\rm DB}$ channel-coupling suppression-factor discussed in Ref.\,\cite[Sec.\,II.E]{Chen:2017pse}.

Focusing on the electric charge $e_\Delta = +2$ state without loss of generality, the Faddeev equation for a $(\tfrac{3}{2},\tfrac{3}{2}^\pm)$ baryon can be written \cite[Sec.\,2.1]{Alkofer:2004yf}, \cite[Sec.\,4.1]{Roberts:2011cf}:
\begin{align}
\sum_{{\mathpzc p}=\pm}&\psi_\lambda^{{\mathpzc p}\pm}(k;Q) \nonumber \\
& =
8\sum_{{\mathpzc p}=\pm} \int\frac{d^4\ell}{(2\pi)^4}
{\mathpzc M}^{\mathpzc p}_{\lambda\mu}(k,\ell;Q)
\psi_\mu^{{\mathpzc p}\pm}(\ell;Q) \,.
\label{KernelFaddeev}
\end{align}
Here, $Q^2 = \hat Q^2 M^2 = - M^2$, $M$ is the baryon's mass,
{\allowdisplaybreaks
\begin{subequations}
\label{decomposition}
\begin{align}
\psi_\lambda^\pm(k;Q) & = \sum_{{\mathpzc p}=\pm} \psi_\lambda^{{\mathpzc p}\pm}(k;Q) \,, \\
\psi_\lambda^{{\mathpzc p}\pm}(k;Q) & =  {\mathpzc T}^{{\mathpzc p} \pm}_{\lambda\rho}(k;Q) u_\rho(Q;r) \,, \label{RSspinor}\\
{\mathpzc T}^{+ \pm}_{\lambda\rho}(k;Q) & =
\sum_{i=1}^{8} {\mathpzc v}_+^i(k^2,k\cdot Q) {\mathpzc G}^\pm {\cal V}^i_{\lambda\rho}(k;Q) ,\\
{\mathpzc T}^{- \pm}_{\lambda\rho}(k;Q) &=
\sum_{i=1}^{8} {\mathpzc v}_-^i(k^2,k\cdot Q) {\mathpzc G}^\mp {\cal V}^i_{\lambda\rho}(k;Q),
\end{align}
\end{subequations}
where ${\mathpzc G}^{+(-)} = {\mathbb I}_{\rm D} (i\gamma_5)$ and, with
$T_{\mu\nu}=\delta_{\mu\nu}+\hat Q_\mu \hat Q_\nu$,
$\gamma_\mu^\perp = T_{\mu\nu} \gamma_\nu$,
$k_\mu^\perp = T_{\mu\nu} k_\nu$,
$\hat k_\mu^\perp \hat k_\mu^\perp  = 1$,
\begin{subequations}
\label{AngularMomentum}
\begin{align}
{\cal V}^1_{\lambda\rho}(k;Q) & = \delta_{\lambda\rho}{\mathbb I}_{\rm D}\,,\\
{\cal V}^2_{\lambda\rho}(k;Q) & =
\tfrac{i}{\surd 5}[2 \gamma_\lambda^\perp \hat k^\perp_\rho - 3 \delta_{\lambda\rho}
\gamma\cdot \hat k^\perp]\,,\\
{\cal V}^3_{\lambda\rho}(k;Q) & = -i\gamma^\perp_\lambda \hat k^\perp_\rho\,, \\
{\cal V}^4_{\lambda\rho}(k;Q) & = \surd 3 \hat Q_\lambda \hat k_\rho^\perp \,,\\
{\cal V}^5_{\lambda\rho}(k;Q) & = 3 \hat k^\perp_\lambda \hat k^\perp_\rho - \delta_{\lambda\rho} - \gamma^\perp_\lambda \hat k^\perp_\rho \gamma\cdot\hat k^\perp\,,\\
{\cal V}^6_{\lambda\rho}(k;Q) & = \gamma_\lambda^\perp \hat k_\rho^\perp \gamma\cdot\hat k^\perp\,, \\
{\cal V}^7_{\lambda\rho}(k;Q) & = -i\surd 3 \hat Q_\lambda \hat k_\rho^\perp \gamma\cdot \hat k^\perp\,, \\
{\cal V}^8_{\lambda\rho}(k;Q) & = \tfrac{i}{\surd 5} [
\delta_{\lambda \rho} \gamma\cdot \hat k^\perp + \gamma^\perp_\lambda \hat k^\perp_\rho -5 \hat k^\perp_\lambda \hat k^\perp_\rho \gamma\cdot \hat k^\perp]\,.
\end{align}
\end{subequations}}

In Eq.\,\eqref{RSspinor}, $u_\rho(Q;r)$ is a Rarita-Schwinger spinor:
\begin{align}
\frac{1}{2 M} \sum_{r=-3/2}^{3/2} u_\mu(Q;r) \bar u_\nu(Q;r)
= \Lambda_+(Q) R_{\mu\nu}\,,
\end{align}
$\Lambda_+(Q) = (-i\gamma\cdot Q + M)/(2 M)$,
\begin{align}
R_{\mu\nu}& =
\delta_{\mu\nu} {\mathbb I}_{\rm D}  \nonumber \\
& \quad -\tfrac{1}{3}\gamma_\mu \gamma_\nu
+\tfrac{2}{3} \hat Q_\mu \hat Q_\nu {\mathbb I}_{\rm D}
- \tfrac{i}{3} [ \hat Q_\mu \gamma_\nu - \hat Q_\nu \gamma_\mu]\,.
\end{align}
(Details of our Euclidean metric conventions are presented elsewhere \cite[Appendix\,B]{Segovia:2014aza}.)

The kernel in Eq.\,\eqref{KernelFaddeev} can now be constructed from Fig.\,\ref{FigFaddeev}, \emph{e.g}., following the pattern in Ref.\,\cite[Sec.\,4.1]{Roberts:2011cf}:
\begin{align}
{\mathpzc M}_{\lambda\mu}^{\pm} & =
\Gamma_{\sigma}^{1^{\pm}}(k_q-\ell_{qq}/2;\ell_{qq}) S^{T}(\ell_{qq}-k_q) \nonumber \\
& \quad \times
\bar{\Gamma}_{\lambda}^{1^{\pm}}(\ell_{q}-k_{qq}/2;-k_{qq}) S(\ell_q)
D^{1^{\pm}}_{\sigma\mu}(\ell_{qq})\,,\label{RealKernel}
\end{align}
where $\ell_q=\ell+Q/3$, $k_q=k+Q/3$, $\ell_{qq}=-\ell+ 2Q/3$,
$k_{qq}=-k+2Q/3$, and ``T'' denotes matrix transpose.


\begin{table}[t]
\caption{\label{TabL}
Working with the wave function defined in Eq.\,\eqref{FaddeevWF}, decomposed over the basis matrix-vectors in Eq.\eqref{AngularMomentum}, with coefficient functions $\{{\mathpzc w}_\pm^i|i=1,\ldots,8\}$, and projected into the rest frame, one has the tabulated $J=\tfrac{3}{2}=L+S$ angular momentum decomposition.  The last row lists the associated spectroscopic label, with the $J=\tfrac{3}{2}$ subscript suppressed.
 }
\begin{center}
\begin{tabular*}
{\hsize}
{
l@{\extracolsep{0ptplus1fil}}
c@{\extracolsep{0ptplus1fil}}
c@{\extracolsep{0ptplus1fil}}
c@{\extracolsep{0ptplus1fil}}
c@{\extracolsep{0ptplus1fil}}
c@{\extracolsep{0ptplus1fil}}
c@{\extracolsep{0ptplus1fil}}}\hline
$L$ & $0$ & $1$ & $1$ & $2$ & $2$ &$3$ \\
$S$ & $\tfrac{3}{2}$ & $\tfrac{3}{2}$ & $\tfrac{1}{2}$ & $\tfrac{3}{2}$ & $\tfrac{1}{2}$ &$\tfrac{3}{2}\ $ \\[1ex] \hline
$\Psi^{\mathpzc p=\pm}\ $ & ${\mathpzc w}_\pm^1\ $ & ${\mathpzc w}_\pm^{2}\ $ & ${\mathpzc w}_\pm^{3,4}\ $  & ${\mathpzc w}_\pm^{5}\ $ & ${\mathpzc w}_\pm^{6,7}\ $ & ${\mathpzc w}_\pm^8\ $\\
& $^4{\mathsf S}$ & $^4{\mathsf P}$ & $^2{\mathsf P}$ & $^4{\mathsf D}$ & $^2{\mathsf D}$ & $^4{\mathsf F}$ \\\hline
\end{tabular*}
\end{center}
\end{table}

The $(1,1^\pm)$ correlation amplitudes are explained in Ref.\,\cite[Eq.\,(1)]{Chen:2017pse}, but it is useful to recapitulate:
\begin{subequations}
\label{qqBSAs}
\begin{align}
{\Gamma}_\mu^{1^+}(k;K)
    & = i g_{1^+} \gamma_\mu C \, {\mathpzc F}(k^2/\omega_{1^+}^2)\,,\\
{\Gamma}_\mu^{1^-}(k;K)
    & = i g_{{1}^-}  [\gamma_\mu, \gamma\cdot \hat K] \gamma_5 C \, {\mathpzc F}(k^2/\omega_{1^-}^2) \,,
\end{align}
\end{subequations}
where $C=\gamma_2\gamma_4$ is the charge conjugation matrix, ${\cal F}(z)$ is given in Eq.\,\eqref{defcalF}, and the correlation widths are defined by the related masses \cite[Eq.\,(5)]{Chen:2017pse}: $\omega_{1^\pm}^2 = m_{1^\pm}^2/2$.  (The colour and flavour structure has already been absorbed into Eq.\,\eqref{RealKernel}.)  The amplitudes are canonically normalised \cite[Eq.\,(3)]{Chen:2017pse}, which entails:
\begin{equation}
g_{1^+} = 12.7\,,\;
g_{1^-} = 1.58\,.
\end{equation}
Since it is the coupling-squared which appears in the Faddeev kernel, $(1,1^+)$ diquarks should be the overwhelmingly favoured correlations in all states considered herein.
This fact lends support to baryon spectrum calculations made using a symmetry-preserving regularisation of a vector\,$\times$\,vector contact interaction, which cannot support $(1,1^-)$ diquarks \cite{Yin:2021uom, Gutierrez-Guerrero:2021rsx}.

Using the information above, the masses and Faddeev amplitudes of the ground- and first-excited state in both the positive- and negative-parity channels can be obtained straightforwardly by solving the Faddeev equation -- Fig.\,\ref{FigFaddeev},  Eq.\,\eqref{KernelFaddeev} -- using readily available software \cite{Arpack, SPECTRA}.

Importantly for what follows in connection with angular momentum decompositions of baryon properties, the unamputated Faddeev wave function is recovered from the amplitude by reattaching the quark and diquark propagator legs:
{\allowdisplaybreaks
\begin{subequations}
\label{FaddeevWF}
\begin{align}
\Psi^\pm_\lambda(k;Q) & =\sum_{{\mathpzc p}=\pm} \Psi^{{\mathpzc p}\pm}_\lambda(k;Q) \\
& =\sum_{{\mathpzc p}=\pm}
S(k_q) D_{\lambda \mu}^{1^{\mathpzc p}}(k_d)
\psi_\mu^{{\mathpzc p}\pm}(k;Q) \,.
\end{align}
\end{subequations}
It is only when working with the wave function that meaningful angular momentum decompositions become available.
It is straightforward to reformulate the Faddeev equation such that the wave function is returned as the solution eigenvector instead of the amplitude.
}

Decomposing $\Psi^\pm_\lambda(k;Q) $ over the basis matrix-vectors in Eq.\eqref{AngularMomentum}, following the pattern in Eq.\,\eqref{decomposition} but with distinct coefficient functions, written herein as $\{{\mathpzc w}_\pm^i|i=1,\ldots,8\}$, then one has the angular momentum associations listed in Table~\ref{TabL}.

\begin{table}[t]
\caption{\label{DeltaMasses}
Calculated masses of lowest-lying $(\tfrac{3}{2},\tfrac{3}{2}^\pm)$ $\Delta$-baryons:
the indicated uncertainty stems from a $\pm5$\% change in the $(1,1^\pm)$ diquark masses in Eq.\,\eqref{diquarkmasses}.
The mean difference between central predicted masses and the real-part of the empirical pole positions is $\delta_{MB}=0.17\,$GeV.
The remaining columns display the mass fractions contributed by the $(1,1^\pm)$ diquarks, described in connection with Eq.\,\eqref{massfraction}, and analogous amplitude fractions, with the latter defined via Eqs.\,\eqref{AmplitudeFractionqq}.
 }
\begin{center}
\begin{tabular*}
{\hsize}
{
l@{\extracolsep{0ptplus1fil}}
|c@{\extracolsep{0ptplus1fil}}
c@{\extracolsep{0ptplus1fil}}
|c@{\extracolsep{0ptplus1fil}}
c@{\extracolsep{0ptplus1fil}}
|c@{\extracolsep{0ptplus1fil}}
c@{\extracolsep{0ptplus1fil}}}\hline
			\multirow{2}{*}{} & \multicolumn{2}{c|}{mass/GeV} & \multicolumn{2}{c|}{mass \% } & \multicolumn{2}{c}{amplitude \%} \\
			\cline{2-7}
			& $1^+\ $ & $1^+$\&$1^-\ $ & $1^+\ $ & $1^-\ $ & $1^+\ $ & $1^-\ $ \\
			\hline
			$ \Delta(1232)\tfrac{3}{2}^+\ $ & 1.346 & 1.346(89)$\ $ & 99.98 & 0.02$\ $ & 96.97 & \phantom{1}3.03$\ $ \\
			\hline
			$ \Delta(1600)\tfrac{3}{2}^+\ $ & 1.786 & 1.786(79)$\ $ & 99.96 & 0.04$\ $ & 96.57 & \phantom{1}3.43$\ $ \\
			\hline
			$ \Delta(1700)\tfrac{3}{2}^-\ $ & 1.872 & 1.871(69)$\ $ & 99.98 & 0.02$\ $ & 94.20 & \phantom{1}5.80$\ $ \\
			\hline
			$ \Delta(1940)\tfrac{3}{2}^-\ $ & 2.030 & 2.043(50)$\ $ & 99.37 & 0.63$\ $ & 88.73 & 11.27$\ $ \\
			\hline
		\end{tabular*}
\end{center}
\end{table}

\section{Solutions and their features}
\label{Solutions}
\subsection{Quark core}
\label{SecQuarkCore}
\begin{figure}[t]
\centerline{%
\includegraphics[clip, width=0.45\textwidth]{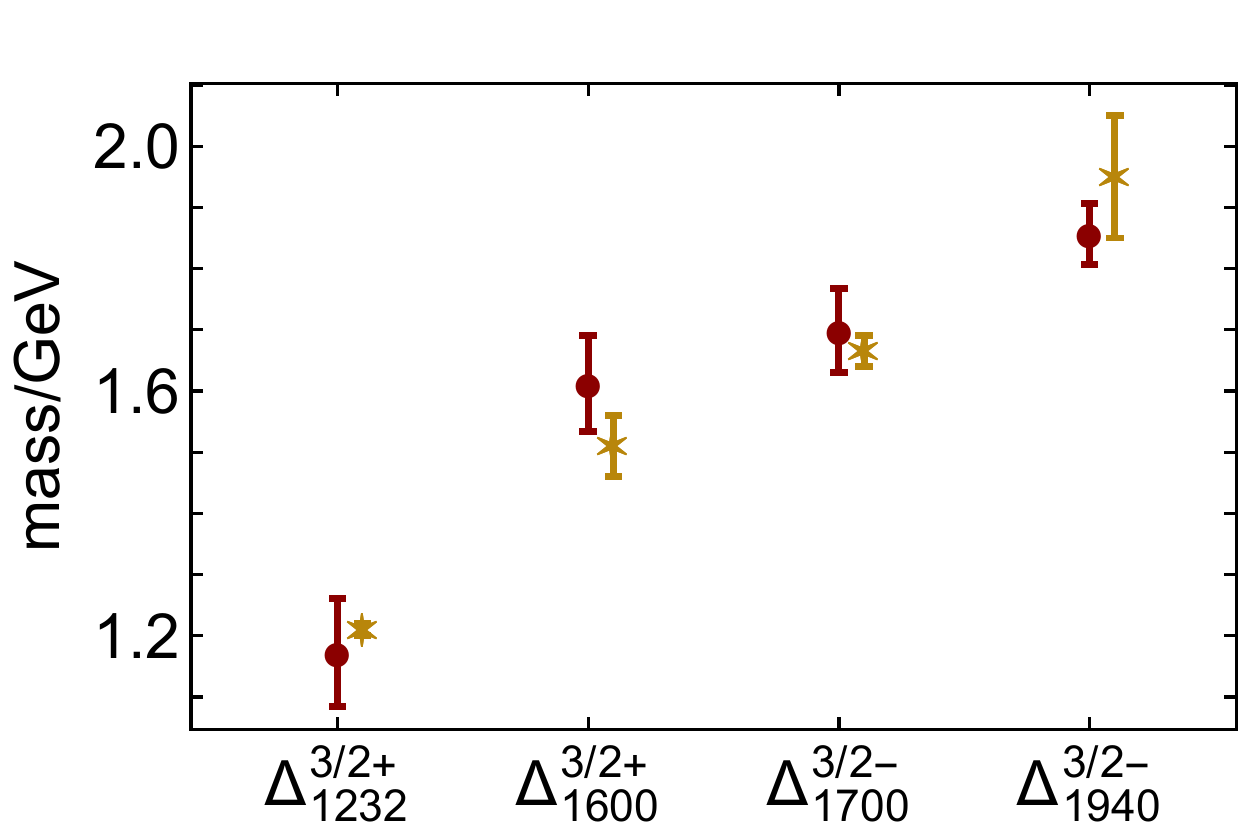}}
\caption{\label{MassCompare}
Real part of empirical pole position for each identified baryon \cite{Zyla:2020zbs} (gold asterisk) compared with calculated masses in Table~\ref{DeltaMasses} after subtracting $\delta_{\rm MB}=0.17\,$GeV from each of the latter (red circles).  The calculated values are drawn with an uncertainty stemming from a $\pm5$\% change in the $(1,1^\pm)$ diquark masses.
}
\end{figure}

Solving for the complete Faddeev amplitude, one obtains the masses listed in Table~\ref{DeltaMasses}.
Notably, the kernel in Fig.\,\ref{FigFaddeev} omits all those contributions which may be linked with meson-baryon final-state interactions, \emph{viz}.\ the terms resummed in dynamical coupled channels (DCC) models in order to transform a bare-baryon into the observed state \cite{JuliaDiaz:2007kz, Suzuki:2009nj, Ronchen:2012eg, Kamano:2013iva}.  Our Faddeev amplitudes should thus be viewed as describing the \emph{dressed-quark core} of the bound-state, not the completely-dressed, observable object \cite{Eichmann:2008ae, Eichmann:2008ef, Roberts:2011cf}; hence, the masses are uniformly too large.  For comparison with experiment, we subtract the mean value of the difference between our calculated masses and the real part of the related empirical pole-positions: $\delta_{\rm MB}=0.17\,$GeV.  This value matches the offset between bare and dressed $\Delta(1232)\tfrac{3}{2}^+$ masses determined in the DCC analysis of Ref.\,\cite{Suzuki:2009nj}.
The resulting comparison is displayed in Fig.\,\ref{MassCompare}: the calculated level orderings and splittings match well with experiment.

The diquark mass fractions in Table~\ref{DeltaMasses} are obtained as follows.
(\emph{i}) Solve for the baryon mass without $(1,1^-)$ diquarks to obtain $m_{\Delta}^{1^+}$.
(\emph{ii}) Solve with both diquarks included to obtain $m_{\Delta}^{1^\pm}$.
(\emph{iii}) The listed fractions are
\begin{equation}
\label{massfraction}
{\rm mass}^{1^+}=m_{\Delta}^{1^+}/m_{\Delta}^{1^\pm}\,,\;
{\rm mass}^{1^-}=1-{\rm mass}^{1^+}.
\end{equation}

Considering the Poincar\'e-covariant Faddeev wave functions obtained for each state, it is worth recording some remarks about the zeroth Chebyshev projection of each term in the wave function analogue of Eq.\,\eqref{decomposition}, expressed using $ {\mathpzc w}^j \in \{{\mathpzc w}_+^{i,\ldots,8}\}\cup \{{\mathpzc w}_-^{i,\ldots,8}\}$:
\begin{equation}
\label{Chebyshev}
{\mathpzc w}^j(k^2) = \frac{2}{\pi}\int_{-1}^1 dx\,\sqrt{1-x^2}\,{\mathpzc w}^j(k^2,x \sqrt{k^2 Q^2})\,.
\end{equation}

The positive-parity states are straightforward: $\Delta(1232)\tfrac{3}{2}^+$ -- no such function with significant magnitude possesses a zero, an outcome consistent with the picture of this system as a radial ground state; and
$\Delta(1600)\tfrac{3}{2}^+$ -- every function with significant magnitude displays a single zero; hence, as explained in connection with meson radial excitations \cite{Holl:2004fr, Li:2016dzv}, this state has the appearance of the radial excitation of the $\Delta(1232)\tfrac{3}{2}^\pm$.
These features are illustrated in Figs.\,\hyperref[NewFigureDeltaPositiveS]{A.1.A}, \hyperref[NewFigureDeltaPositiveS]{A.1.B}.

On the other hand, as found with $(\tfrac{1}{2},\tfrac{1}{2}^-)$ states \cite{Chen:2017pse},
the wave functions of the negative-parity $\Delta$-baryons are much more complex.
This is illustrated in 
Figs.\,\hyperref[NewFigureDeltaPositiveS]{A.1.C}, \hyperref[NewFigureDeltaPositiveS]{A.1.D},
which show that for both $\Delta(1700)\tfrac{3}{2}^-$ and $\Delta(1940)\tfrac{3}{2}^-$ most of the wave function projections, Eq.\,\eqref{Chebyshev}, possess a zero; and this is true for more of the $\Delta(1940)\tfrac{3}{2}^-$ projections.  

It is worth noting that when a zero exists, it lies within the domain $\tfrac{1}{3}{\rm fm}\lesssim \frac{1}{k} \lesssim \tfrac{1}{2}{\rm fm}$, \emph{i.e}., at length-scales smaller than the bound-state radii.  This is similarly so of $(\tfrac{1}{2},\tfrac{1}{2}^\pm)$ bound-states \cite[Figs.\,4, 5]{Chen:2017pse} and also vector mesons \cite[Fig.\,5]{Qin:2011xq}.  The zero in the leading Chebyshev amplitude of the pion's first radial excitation is found even deeper: $\tfrac{1}{k} \approx \tfrac{1}{5}\,$fm \cite[Fig.\,4]{Qin:2011xq}.

Such structural predictions for the properties of $(\tfrac{3}{2},\tfrac{3}{2}^\pm)$ baryons can be tested via comparisons with data obtained on the $Q^2$-dependence of nucleon-to-resonance transition form factors \cite{Brodsky:2020vco, Carman:2020qmb, Mokeev:2022xfo}.

\subsection{Diquark fractions}
\label{qqFractions}
It is apparent from Table~\ref{DeltaMasses} that, so far as the masses are concerned, neglecting $(1,1^-)$ diquark correlations is an excellent approximation.  One can also consider their relative contribution to the Faddeev amplitude, which may be defined following Ref.\,\cite{Chen:2017pse}.  Namely, with
\begin{equation}
		{\mathpzc n}^j = \int \dfrac{d^4k}{(2\pi)^4} |{\mathpzc u}^j(k^2, k \cdot Q)|^2 \,,
\end{equation}
where $ {\mathpzc u}^j \in \{{\mathpzc v}_+^{i,\ldots,8}\}\cup \{{\mathpzc v}_-^{i,\ldots,8}\}$, then, for each $\Delta$-baryon, one computes
\begin{equation}
\mathbb{N}_{{\mathpzc p}=\pm} = 
\sum_{j \in \{ {\mathpzc v}_{{\mathpzc p}}^{i,\ldots,8}\}}\!\! {\mathpzc n}^j\,, \quad
{\mathbb D} = \mathbb{N}_+ + \mathbb{N}_-
\label{AmplitudeFractionqq}
\end{equation}
and compares the results for $\mathbb{F}_\pm = \mathbb{N}_\pm/{\mathbb D}$, which are listed, respectively, in the final two columns of Table~\ref{DeltaMasses}.  Unsurprisingly, the negative-parity diquarks feature most prominently in the negative-parity baryons; but even in these states, they are very much subdominant.

\subsection{Angular momentum decompositions}
\label{SecLD}
We judge it to be of particular interest to expose the rest-frame angular momentum structure of the $(\tfrac{3}{2},\tfrac{3}{2}^\pm)$ systems produced by our Poincar\'e covariant framework.  As a first step toward that goal, we solved the Faddeev equation for the wave function of each baryon in its rest frame by changing and steadily increasing the orbital angular momentum complexity:
(\emph{i}) ${\mathsf S}$-wave only; (\emph{ii}) ${\mathsf P}$-wave only; (\emph{iii}) ${\mathsf D}$-wave only; (\emph{iv}) ${\mathsf S}+{\mathsf P}$-wave only; etc.  The results are presented in Table~\ref{TabSPDetc}.

\begin{figure}[t]
\vspace*{1ex}

\centerline{%
\includegraphics[clip, width=0.45\textwidth]{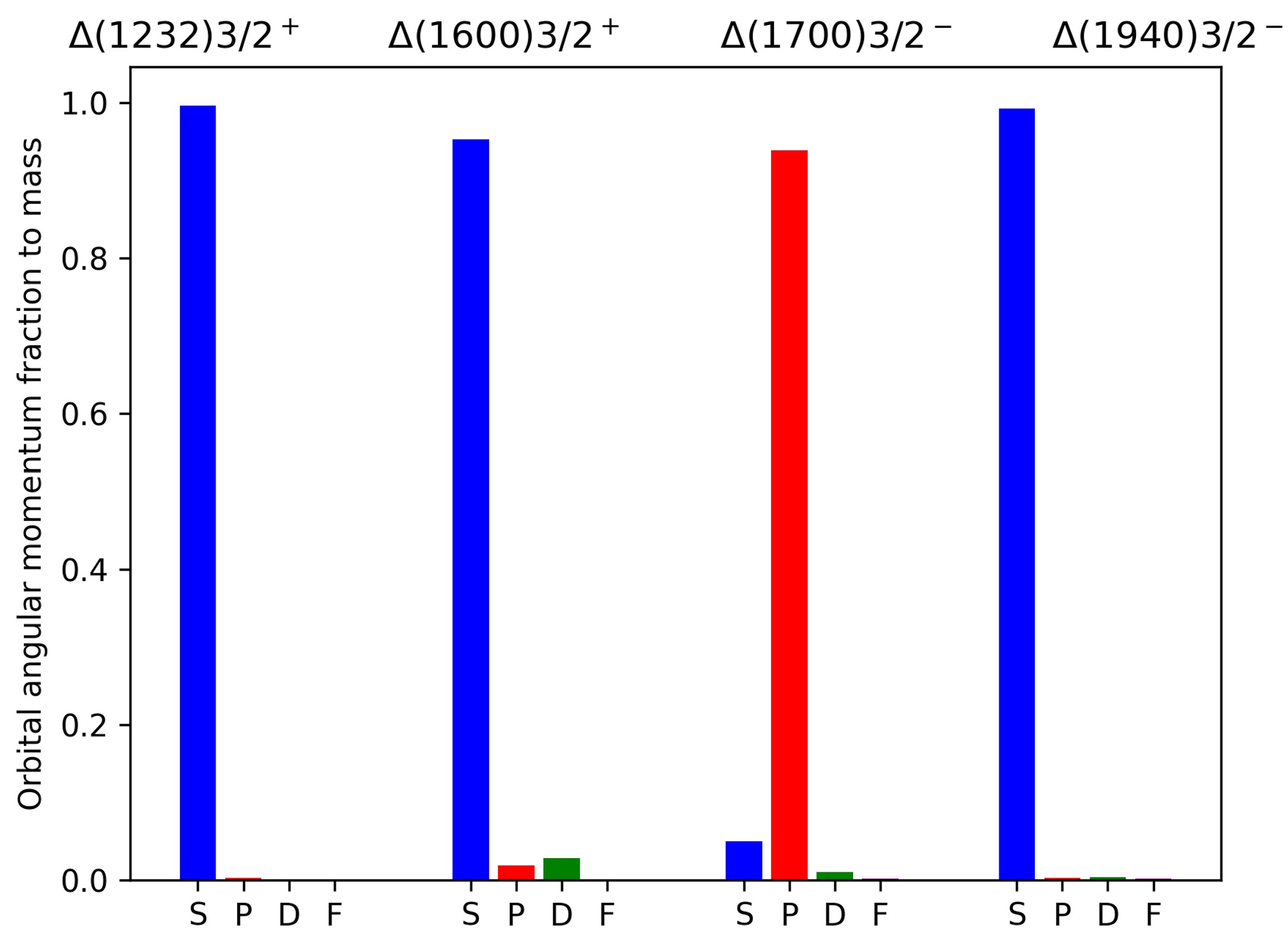}}
\caption{\label{LMass}
Pictorial representation of Table~\ref{TabSPDetc}.
Mass fraction contribution from each rest frame partial wave in the baryon wave function, computed as follows:
$ \Delta(1232)\tfrac{3}{2}^+$, $\Delta(1600)\tfrac{3}{2}^+$, $\Delta(1940)\tfrac{3}{2}^-$ -- begin with $\mathsf S$-wave, then add $\mathsf P$, $\mathsf D$, $\mathsf F$;
and
$\Delta(1700)\tfrac{3}{2}^-$ -- begin with $\mathsf P$-wave, then add $\mathsf S$, $\mathsf D$, $\mathsf F$.
}
\end{figure}

Table~\ref{TabSPDetc} rewards careful inspection.
For instance, it reveals that in every channel a solution is obtained using only one partial wave -- $\mathsf S$, $\mathsf P$, $\mathsf D$, or $\mathsf F$, or any subset of the complete array of partial waves.
%
%
Plainly, notwithstanding its apparent simplicity, the Faddeev kernel in Eq.\,\eqref{RealKernel} is very effective at binding $(\tfrac{3}{2},\tfrac{3}{2}^\pm)$ baryons.
%
Furthermore, considering only a single partial wave, then the lightest mass obtained should serve as a reliable indicator of the dominant orbital angular momentum component in the state.  Using this definition, one arrives at the following assignments:
$ \Delta(1232)\tfrac{3}{2}^+$ and $ \Delta(1600)\tfrac{3}{2}^+$ are largely ${\mathsf S}$-wave in nature, but with contributing ${\mathsf P}$-, ${\mathsf D}$-wave components;
$ \Delta(1700)\tfrac{3}{2}^-$ is primarily a ${\mathsf P}$-wave state, but possesses measurable ${\mathsf S}$-, ${\mathsf D}$-wave components;
and, surprisingly, because it runs counter to quark model notions \cite[Sec.\,15]{Zyla:2020zbs}, $\Delta(1940)\tfrac{3}{2}^-$ is predominantly a ${\mathsf S}$-wave state, with small contributions from other partial waves.
These observations are illustrated in Fig.\,\ref{LMass}.

\begin{table}[b]
\caption{\label{TabSPDetc}
Calculated masses of the lowest-lying $(\tfrac{3}{2},\tfrac{3}{2}^\pm)$ $\Delta$-baryons (in GeV) as obtained by stepwise including different orbital angular momentum components in the rest-frame Faddeev wave function.  The italicised entries highlight the lowest mass obtained in solving with a single partial wave.
 }
\begin{center}
\begin{tabular*}
{\hsize}
{
l@{\extracolsep{0ptplus1fil}}
|c@{\extracolsep{0ptplus1fil}}
c@{\extracolsep{0ptplus1fil}}
c@{\extracolsep{0ptplus1fil}}
c@{\extracolsep{0ptplus1fil}}
|c@{\extracolsep{0ptplus1fil}}
c@{\extracolsep{0ptplus1fil}}
c@{\extracolsep{0ptplus1fil}}
|c@{\extracolsep{0ptplus1fil}}
|c@{\extracolsep{0ptplus1fil}}}\hline
	$\Delta$		~ & ${\mathsf S}$ & ${\mathsf P}$ & ${\mathsf D}$ & ${\mathsf F}$ & $\mathsf{SP}$ & $\mathsf{SD}$ & $\mathsf{PD}\ $ & $\mathsf{SPD}$$\ $ & $\mathsf{SPDF}$  \\ \hline
$ (1232)\tfrac{3}{2}^+\ $  & \emph{1.35} & 2.03 & 1.80 & 2.35$\ $ & 1.35 & 1.36$\ $ & 
1.83$\ $ & 1.35$\ $ & 1.35  \\ \hline
$ (1600)\tfrac{3}{2}^+\ $ & \emph{1.80} & 2.22 & 2.10& 2.48$\ $ & 1.84 & 1.76$\ $ & 
2.02$\ $ & 1.78$\ $ & 1.79  \\ \hline
$ (1700)\tfrac{3}{2}^-\ $ & 1.90 & \emph{1.80} & 2.18 & 2.17$\ $ & 1.89 & 1.90$\ $ & 1.80$\ $ & 1.87$\ $ & 1.87  \\ \hline
$ (1940)\tfrac{3}{2}^-\ $ & \emph{2.06} & 2.20 & 2.27& 2.38$\ $ & 2.05 & 2.05$\ $ & 2.19$\ $ & 2.05$\ $ & 2.04  \\ \hline
\end{tabular*}
\end{center}
\end{table}

\begin{figure}[t]
\vspace*{1ex}

\centerline{%
\includegraphics[clip, width=0.4\textwidth]{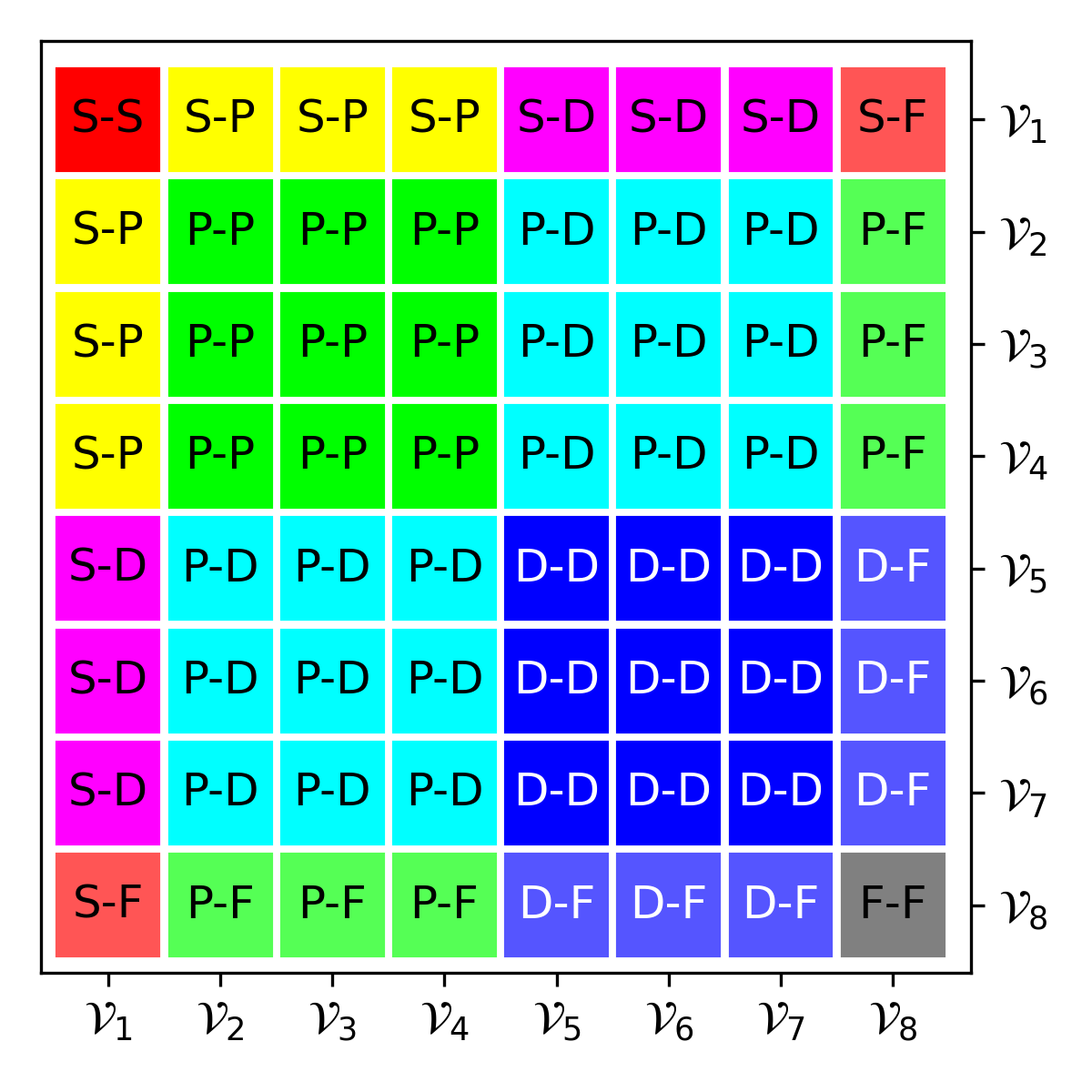}}
\caption{\label{LWFlegend}
Legend for interpretation of Figs.\,\ref{LFigures}A\,--\,D, identifying interference between the various identified orbital angular momentum basis components in the baryon rest frame.
}
\end{figure}

Hadron masses are simple observables in the sense that they are infrared dominated quantities, whose values are not especially sensitive to structural details expressed in hadron wave functions.  Consequently, the simplicity evident in Fig.\,\ref{LMass} is somewhat misleading, as highlighted again when one isolates the distinct contributions from each partial wave to the associated canonical normalisation. 
Using the assignments specified in Fig.\,\ref{LWFlegend}, those decompositions are depicted in Fig.\,\ref{LFigures}, being drawn from the numerical values collected in Appendix~\ref{AAngular}. Since $(1,1^-)$ diquarks make negligible contributions, only the $(1,1^+)$ contributions are reported and drawn.

Considering Fig.\,\ref{LFigures}A, one sees that, evaluated in the rest frame, the canonical normalisation of the $\Delta(1232)\tfrac{3}{2}^+$ is largely determined by $\mathsf S$-wave components, but there are significant, constructive $\mathsf P$ wave contributions and also strong $\mathsf S\otimes \mathsf P$-wave destructive interference terms.  This structural picture of the $\Delta(1232)\tfrac{3}{2}^+$ has been confirmed by comparisons with data on the $\gamma+p \to \Delta(1232)$ transition form factors \cite{Eichmann:2011aa, Segovia:2014aza, Lu:2019bjs}.

\begin{figure*}[!t]
\hspace*{-1ex}\begin{tabular}{lcl}
\large{\textsf{A}} & & \large{\textsf{B}}\\[-0.7ex]
%
\includegraphics[clip, width=0.41\textwidth]{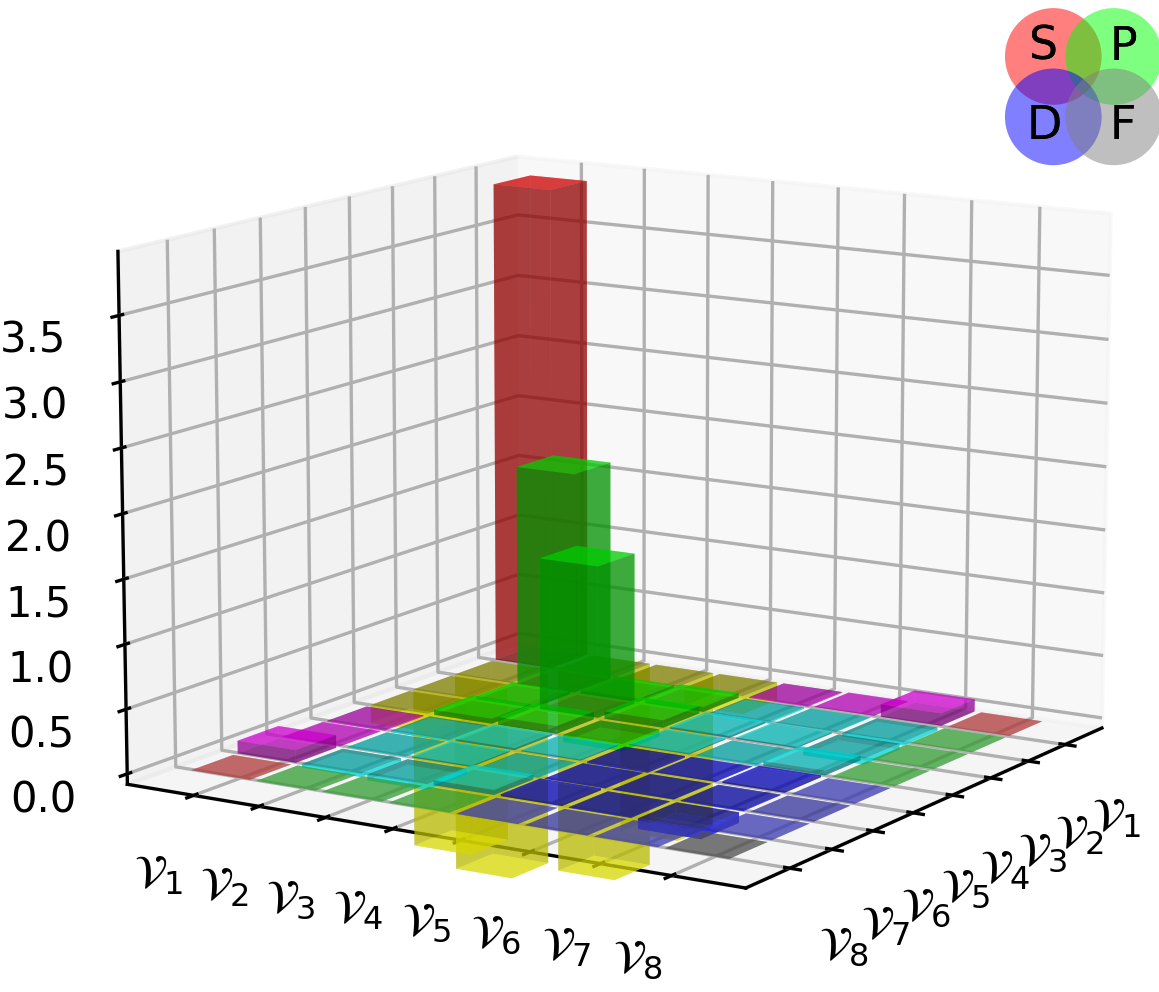} & \hspace*{4em} &
\includegraphics[clip, width=0.41\textwidth]{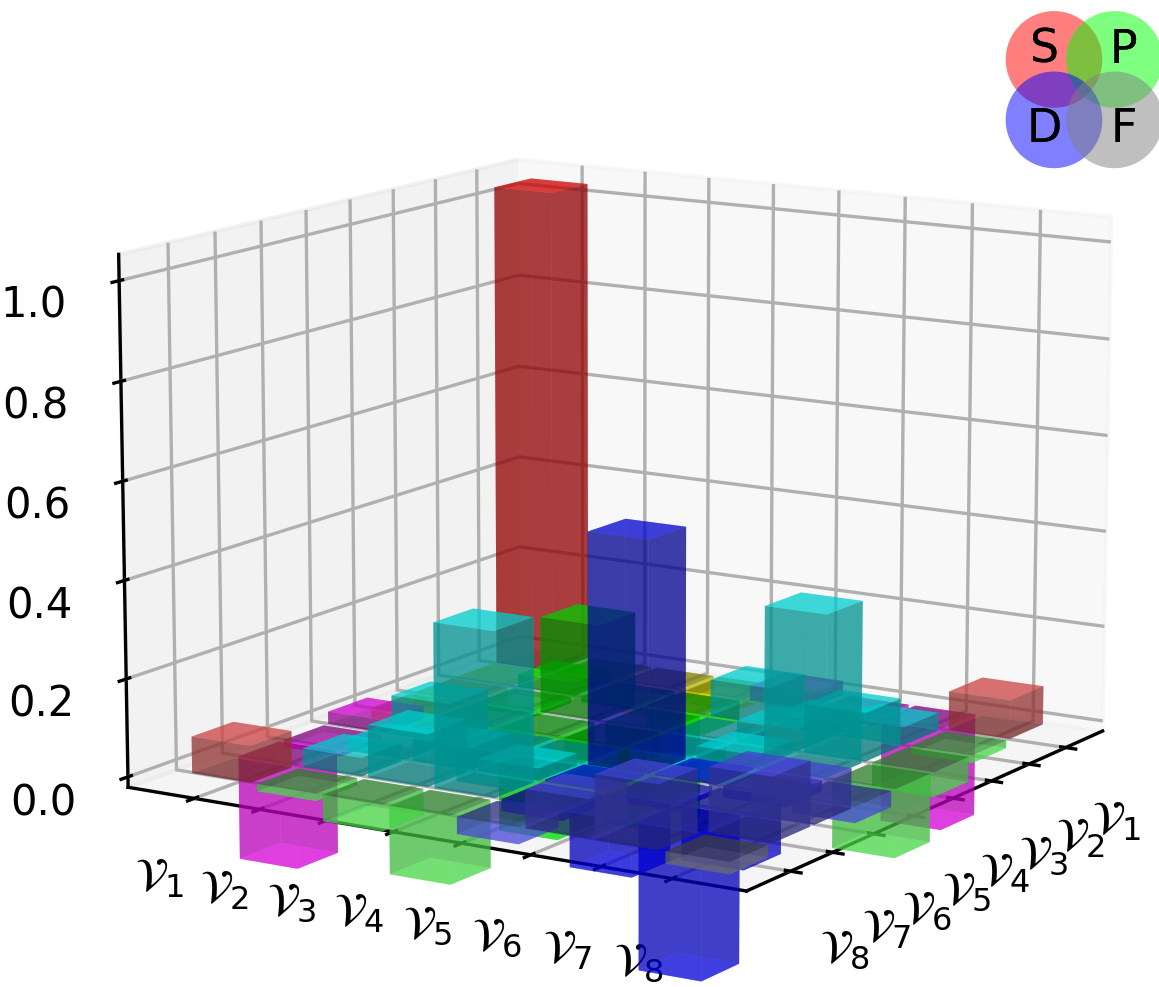} \\
\large{\textsf{C}} & & \large{\textsf{D}}\\[-0.7ex]
\includegraphics[clip, width=0.41\textwidth]{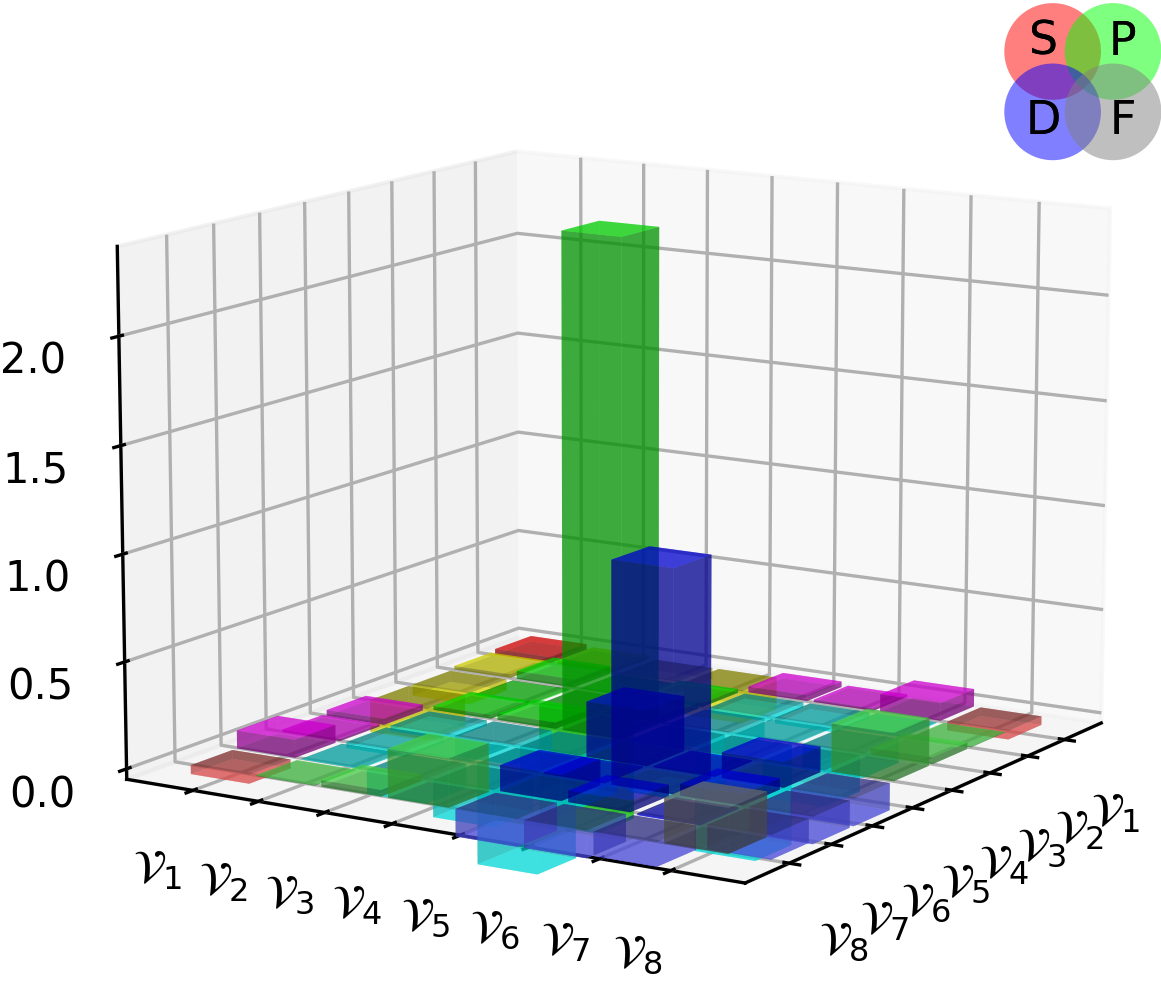} & \hspace*{4em} &
\includegraphics[clip, width=0.41\textwidth]{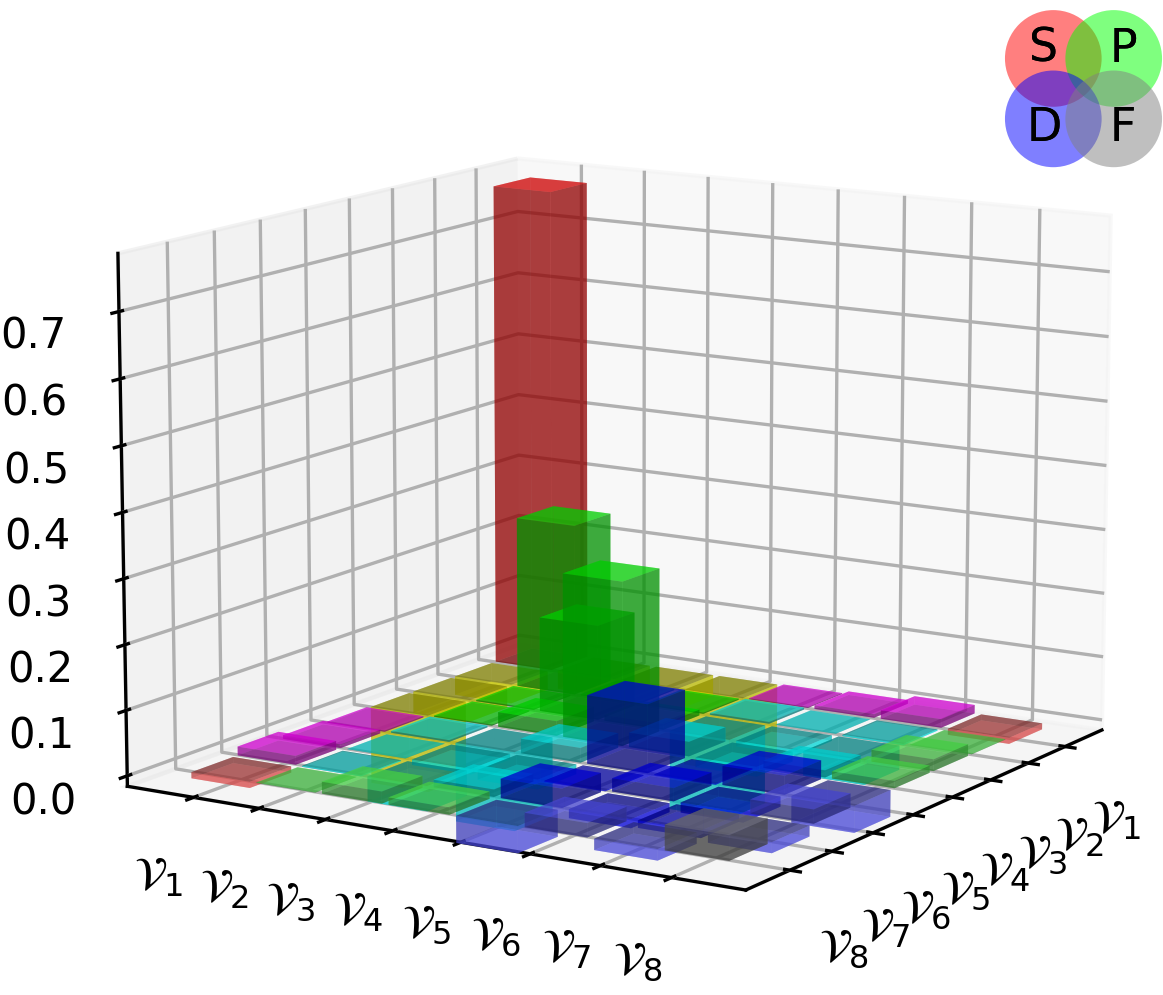} \\
\end{tabular}
\caption{\label{LFigures}
Rest frame quark+$(1,1^+)$-diquark orbital angular momentum content of $(\tfrac{3}{2},\tfrac{3}{2}^\pm)$ states considered herein, as measured by the contribution of the various components to the associated canonical normalisation constant:
{\sf A} -- $ \Delta(1232)\tfrac{3}{2}^+$;
{\sf B} -- $ \Delta(1600)\tfrac{3}{2}^+$;
{\sf C} -- $ \Delta(1700)\tfrac{3}{2}^-$; and
{\sf D} -- $ \Delta(1940)\tfrac{3}{2}^-$ --
drawn with reference to Table~\ref{TabL} and the basis in Eq.\,\eqref{AngularMomentum}.
There are both positive (above plane) and negative (below plane) contributions to the overall normalisations, which are all positive.}
\end{figure*}

Moving to Fig.\,\ref{LFigures}B, although $\mathsf S$-wave contributions are dominant in the $\Delta(1600)\tfrac{3}{2}^+$, there are prominent $\mathsf D$-wave components, material $\mathsf P \otimes \mathsf D$-wave interference contributions, and numerous $\mathsf F$-wave induced interference terms.
Enhanced higher partial waves are also seen in related three-body Faddeev equation studies of the $\Delta(1600)\tfrac{3}{2}^+$ \cite{Eichmann:2016hgl, Qin:2018dqp}.
This quark+diquark structural picture of the $\Delta(1600)\tfrac{3}{2}^+$ has been used to calculate $\gamma+p \to \Delta(1600)$ transition form factors \cite{Lu:2019bjs}.  Those predictions are currently being tested through analysis of $\pi^+ \pi^- p$ electroproduction data collected at Jefferson Lab \cite{MokeevPrivate2022}.

The $\Delta(1700)\tfrac{3}{2}^-$ normalisation strengths are displayed in Fig.\,\ref{LFigures}C.  Confirming expectations raised by Table~\ref{TabSPDetc}, $\mathsf P$-wave components are dominant, but $\mathsf D$-wave and $\mathsf P \otimes \mathsf D$ interference is evident, and also some $\mathsf D \otimes \mathsf F$ contributions.  $\Delta(1700)\tfrac{3}{2}^-$ electrocoupling data are available from Jefferson Lab \cite{Burkert:2002zz, CLAS:2009tyz, Mokeev:2013kka}.  However, they only reach $Q^2\approx 1.5 m_p^2$; hence, are insufficient to test our $\Delta(1700)\tfrac{3}{2}^-$ structure predictions.  It would nevertheless be worthwhile to use our wave functions as the basis for calculating the $\gamma+p \to \Delta(1700)$ transition form factors, providing motivation and support
for extraction of $\Delta(1700)\tfrac{3}{2}^-$ electrocouplings on $2<Q^2/{\rm GeV}^2<5$ from existing $\pi^+ \pi^- p$ electroproduction data \cite{CLAS:2017fja, Trivedi:2018rgo}.

The $\Delta(1940)\tfrac{3}{2}^-$ normalisation strengths are displayed in Fig.\,\ref{LFigures}D.  Unlike the other systems studied herein, this is only a ``$\ast\ast$'' state \cite{Zyla:2020zbs}; and no electrocoupling data are available, although they are expected to be collected in future Jefferson Lab $\pi^+ \pi^- p$ electroproduction experiments \cite{MokeevPrivate2022}.  Such data would be valuable because our analysis shows that the $\Delta(1940)\tfrac{3}{2}^-$  is potentially a peculiar system, \emph{viz}.\ a negative-parity baryon whose rest-frame wave function is largely $\mathsf S$-wave in character.  Even if this outcome were to indicate a failure of our Faddeev equation in describing some higher baryon resonances, resolving the question is necessary in order to ensure arrival at a reliable Poincar\'e covariant description of  baryon spectra and structure.

\section{Summary and Perspective}
\label{epilogue}
A Poincar\'e-covariant Faddeev equation [Fig.\,\ref{FigFaddeev}], whose kernel is built using dressed-quark and nonpointlike diquark degrees-of-freedom, with binding generated by the exchange of a dressed-quark, which emerges as one diquark breaks-up and is absorbed into formation of another, was used to calculate the mass and Faddeev wave functions of the lowest-lying $(I,J^P = \tfrac{3}{2},\tfrac{3}{2}^\pm)$ baryons.
This framework has previously been used widely to deliver explanations of many baryon properties \cite{Barabanov:2020jvn}, with recent applications to
parton distribution functions \cite{Chang:2022jri, Lu:2022cjx},
the large-$Q^2$ behaviour of elastic form factors \cite{Cui:2020rmu},
and axial form factors \cite{Chen:2020wuq, Chen:2021guo}.
It should, therefore, provide a sound approach to the study of $\Delta$-baryons.

In principle, viewed from the quark+diquark perspective, $(\tfrac{3}{2},\tfrac{3}{2}^\pm)$ baryons can contain both $(1,1^+)$ and $(1,1^-)$ quark+quark correlations.  However, our analysis revealed that $(1,1^-)$ diquarks may safely be neglected [Sec.\,\ref{qqFractions}].
In this case, the Poincar\'e-covariant wave functions of $(\tfrac{3}{2},\tfrac{3}{2}^\pm)$ systems contain eight independent terms, each characterised by a scalar function of two variables: $k^2$, $k\cdot Q$, where $k$ is the quark+diquark relative momentum and $Q$ is the bound-state total momentum.  Projecting each of these functions to obtain their zeroth Chebyshev moment, one arrives at a collection of simpler functions, useful for developing insights.   Reviewing their behaviour [Sec.\,\ref{SecQuarkCore}], we found that the $\Delta(1600)\tfrac{3}{2}^+$ exhibits characteristics which enable it to be interpreted as a radial excitation of the $\Delta(1232)\tfrac{3}{2}^+$.  However, no such simple relationship was found to be viable for the $\Delta(1700)\tfrac{3}{2}^-$, $\Delta(1940)\tfrac{3}{2}^-$ states.

Although the $J=L+S$ separation of a baryon's total angular momentum into a sum of orbital angular momentum and spin is frame dependent, one may nevertheless make some contact with quark model pictures of $(\tfrac{3}{2},\tfrac{3}{2}^\pm)$ baryons by projecting their Poincar\'e-covariant Faddeev wave functions into the associated rest frames.  Following this procedure [Sec.\,\ref{SecLD}], we found that the angular momentum structure of all these $\Delta$-baryons is far more complicated than generated by typical quark models.
Nevertheless, drawing some link to quark models, the $\Delta(1232)\tfrac{3}{2}^+$ and $\Delta(1600)\tfrac{3}{2}^+$ baryons were found to be characterised by a dominant $\mathsf S$-wave component, and the $\Delta(1700)\tfrac{3}{2}^-$ by a prominent $\mathsf P$-wave.  However, the $\Delta(1940)\tfrac{3}{2}^-$ did not fit this picture: contrary to quark model expectations, this state is $\mathsf S$-wave dominated.
Furthermore, combining the results from our analyses of their Poincar\'e-covariant quark+diquark Faddeev wave functions, we judged that negative parity $\Delta$-baryons are not simply orbital angular momentum excitations of positive parity ground states.  This conclusion matches that drawn elsewhere for $(\tfrac{1}{2},\tfrac{1}{2}^\pm)$ baryons \cite{Chen:2017pse}.
Our structural predictions for the $\Delta(1940)\tfrac{3}{2}^-$ are likely to encourage new experimental efforts to extract reliable information about this poorly understood state from exclusive $\pi^+\pi^- p$ electroproduction data \cite{CLAS:2017fja, Trivedi:2018rgo} and subsequent determination of this resonance's electroexcitation amplitudes.


It is here worth recalling that the interpolating fields for positive and negative parity hadrons can be related by chiral rotation of the quark spinors used in their construction.  Hence, all differences between  bound states in these channels are generated by chiral symmetry breaking, which is predominantly dynamical in the light-quark sector.  Regarding the baryons discussed herein, this means that the following states are parity partners: $\Delta(1232)\tfrac{3}{2}^+$\,--\,$\Delta(1700)\tfrac{3}{2}^-$; and $\Delta(1600)\tfrac{3}{2}^+$\,--\,$\Delta(1940)\tfrac{3}{2}^-$.

The mass splitting between parity partners is usually ascribed to dynamical chiral symmetry breaking (DCSB); and we have seen herein that, again like the $(\tfrac{1}{2},\tfrac{1}{2}^\pm)$ sector, there are also marked differences between their internal structures.  They, too, must owe to DCSB because the channels are identical when chiral symmetry is restored.
DCSB is a corollary of emergent hadron mass, which may also be argued to underly confinement \cite{Roberts:2021nhw}; so, validating our predictions of marked structural differences between parity partners has the potential to reveal a great deal about key features of the Standard Model.  A means to this end exists in resonance electroexcitation experiments on $Q^2 \gtrsim 2\,m_p^2$.

There are many natural extensions of this study.  For instance,
solving Faddeev equations to develop insights into the composition of $(\tfrac{1}{2},\tfrac{3}{2}^\pm)$ and $(\tfrac{3}{2},\tfrac{1}{2}^\pm)$ baryons;
calculation of the electromagnetic transition form factors mentioned above and those involving the additional states just indicated;
analyses that focus on the structure of baryons containing heavier valence quarks;
and the prediction of weak proton-to-$\Delta$ transition form factors, which may be crucial in understanding neutrino oscillation experiments \cite{Mosel:2016cwa}.
Efforts are underway in each of these areas.

%
%
\begin{acknowledgments}
We are grateful for constructive comments from Z.-F.~Cui and V.\,I.~Mokeev.
This work was completed using the computer clusters at the Nanjing University Institute for Nonperturbative Physics.
Work supported by:
National Natural Science Foundation of China (grant nos.\,12135007, 12047502);
Jiangsu Province Fund for Postdoctoral Research (grant no.\,2021Z009);
Ministerio Espa\~nol de Ciencia e Innovaci\'on (grant no.\,PID2019-107844GB-C22);
and Junta de Andaluc\'ia (contract nos.\ operativo FEDER Andaluc\'ia 2014-2020 UHU-1264517,  P18-FR-5057, PAIDI FQM-370).
\end{acknowledgments}

\appendix
\setcounter{equation}{0}
\setcounter{figure}{0}
\setcounter{table}{0}
\renewcommand{\theequation}{\Alph{section}.\arabic{equation}}
\renewcommand{\thetable}{\Alph{section}.\arabic{table}}
\renewcommand{\thefigure}{\Alph{section}.\arabic{figure}}

\section{Supplemental Material}
\label{LTables}
\subsection{Quark and diquark propagators}
\label{APropagators}
The dressed-quark propagator can be written:
\begin{align}
S(k)  & =  -i \gamma\cdot k\, \sigma_V(k^2) + \sigma_S(k^2) \\
& = 1/[i\gamma\cdot k\, A(k^2) + B(k^2)]\,.
\label{SpAB}
\end{align}
In QCD, the wave function renormalisation and dressed-quark mass:
\begin{equation}
\label{ZMdef}
Z(k^2)=1/A(k^2)\,,\;M(k^2)=B(k^2)/A(k^2)\,,
\end{equation}
respectively, receive strong momentum-dependent corrections at infrared momenta \cite{Roberts:2021nhw, Oliveira:2018lln}: $Z(k^2)$ is suppressed and $M(k^2)$ enhanced.  These features are an expression of DCSB. 

\begin{figure*}[!t]
\begin{AAfigure}
\label{NewFigureDeltaPositiveS}
\addtocounter{Afigure}{1}
\end{AAfigure}
\hspace*{-1ex}\begin{tabular}{lcl}
\large{\textsf{A}} & & \large{\textsf{B}}\\[-0.7ex]
%
\includegraphics[clip, width=0.4\textwidth]{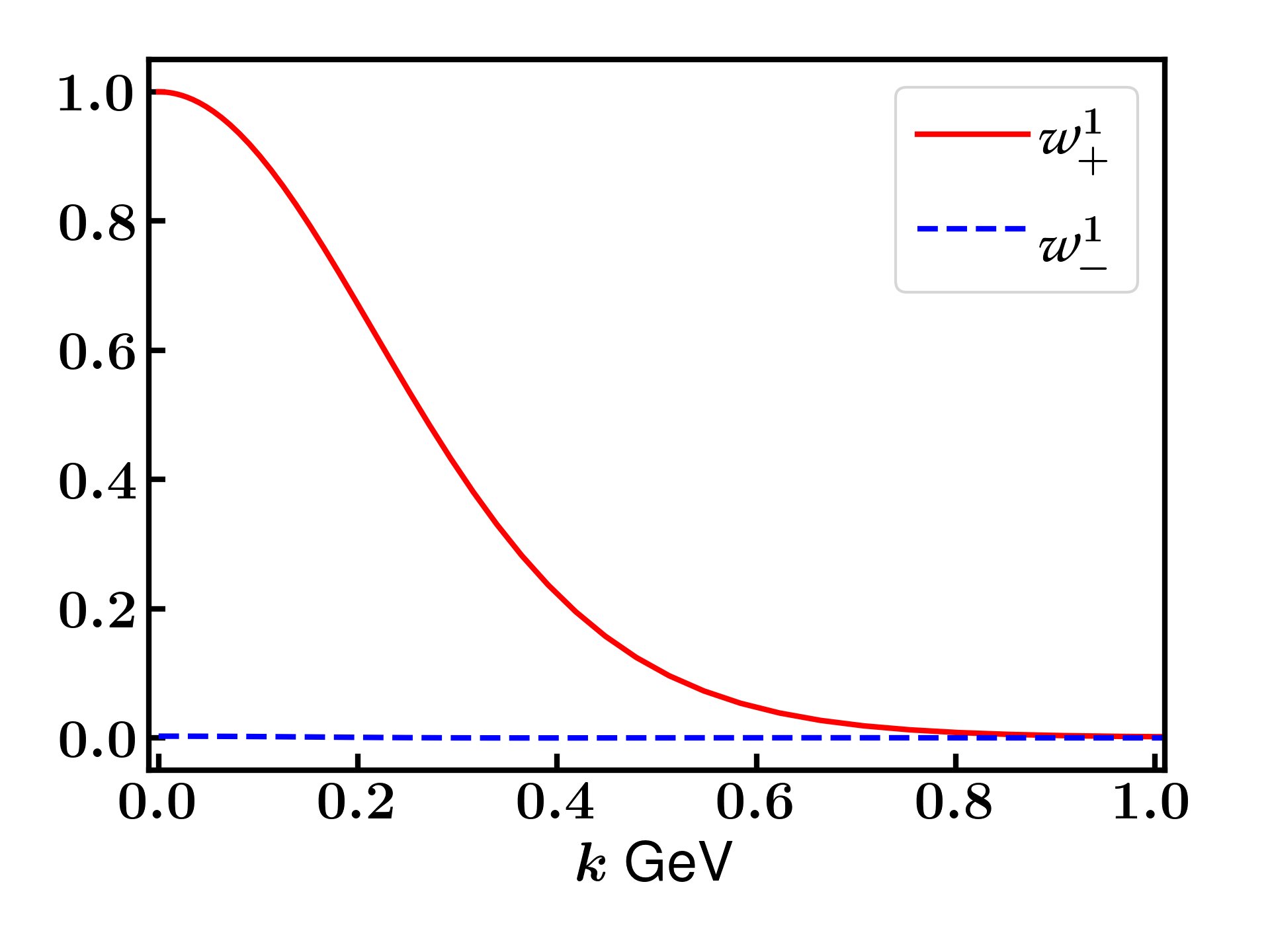} & \hspace*{4em} &
\includegraphics[clip, width=0.4\textwidth]{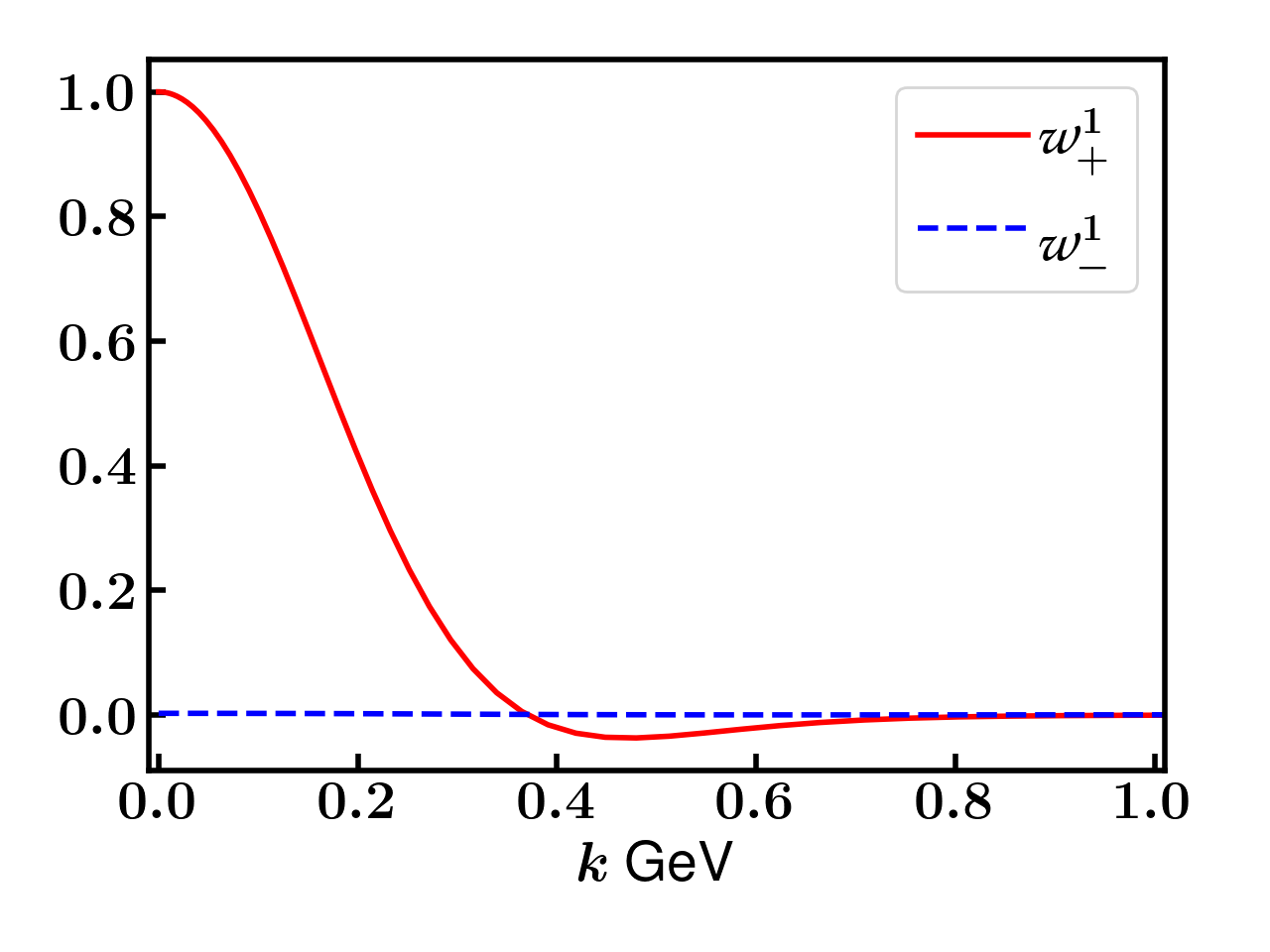} \\
\large{\textsf{C}} & & \large{\textsf{D}}\\[-0.7ex]
\includegraphics[clip, width=0.4\textwidth]{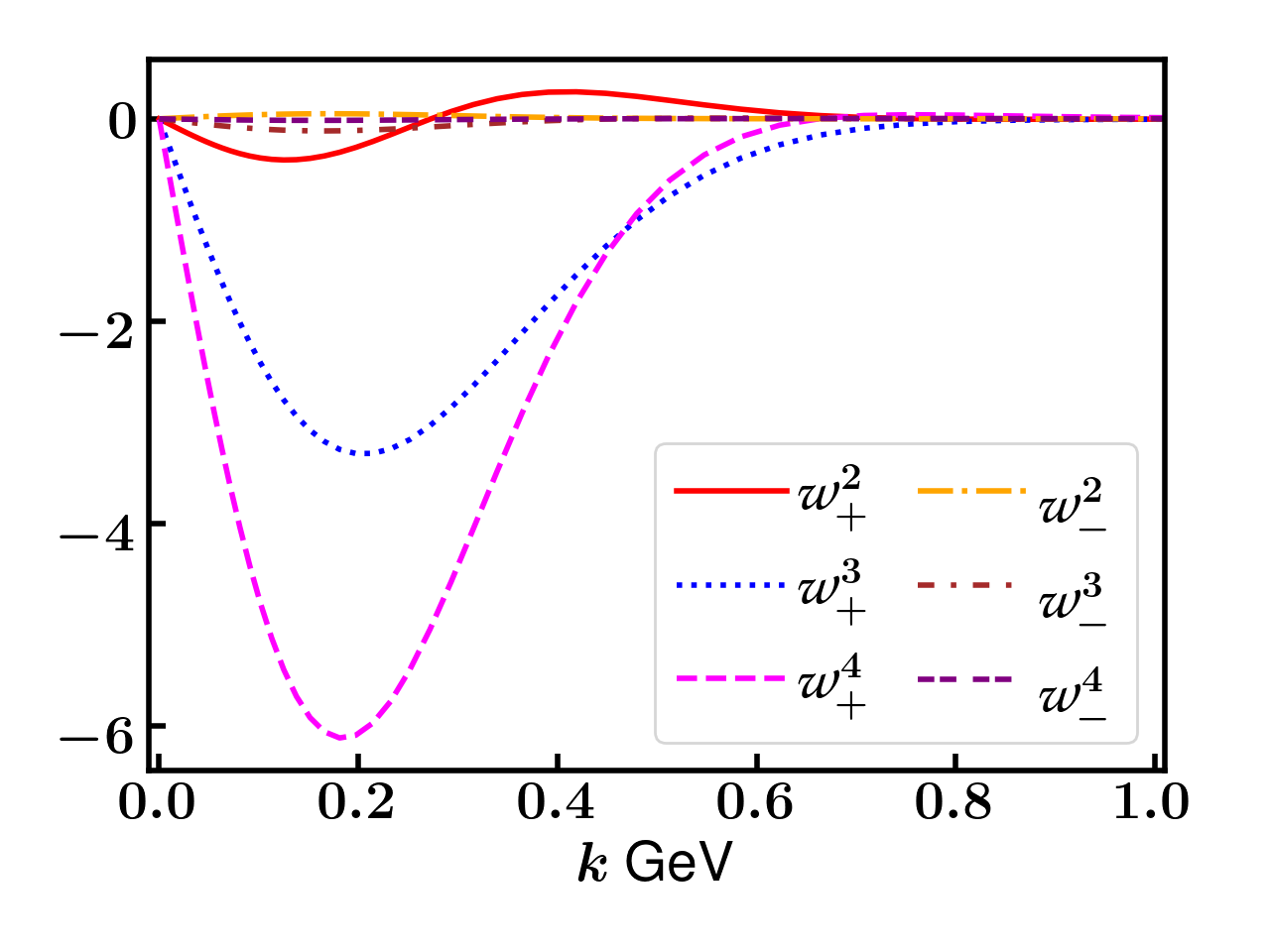} & \hspace*{4em} &
\includegraphics[clip, width=0.4\textwidth]{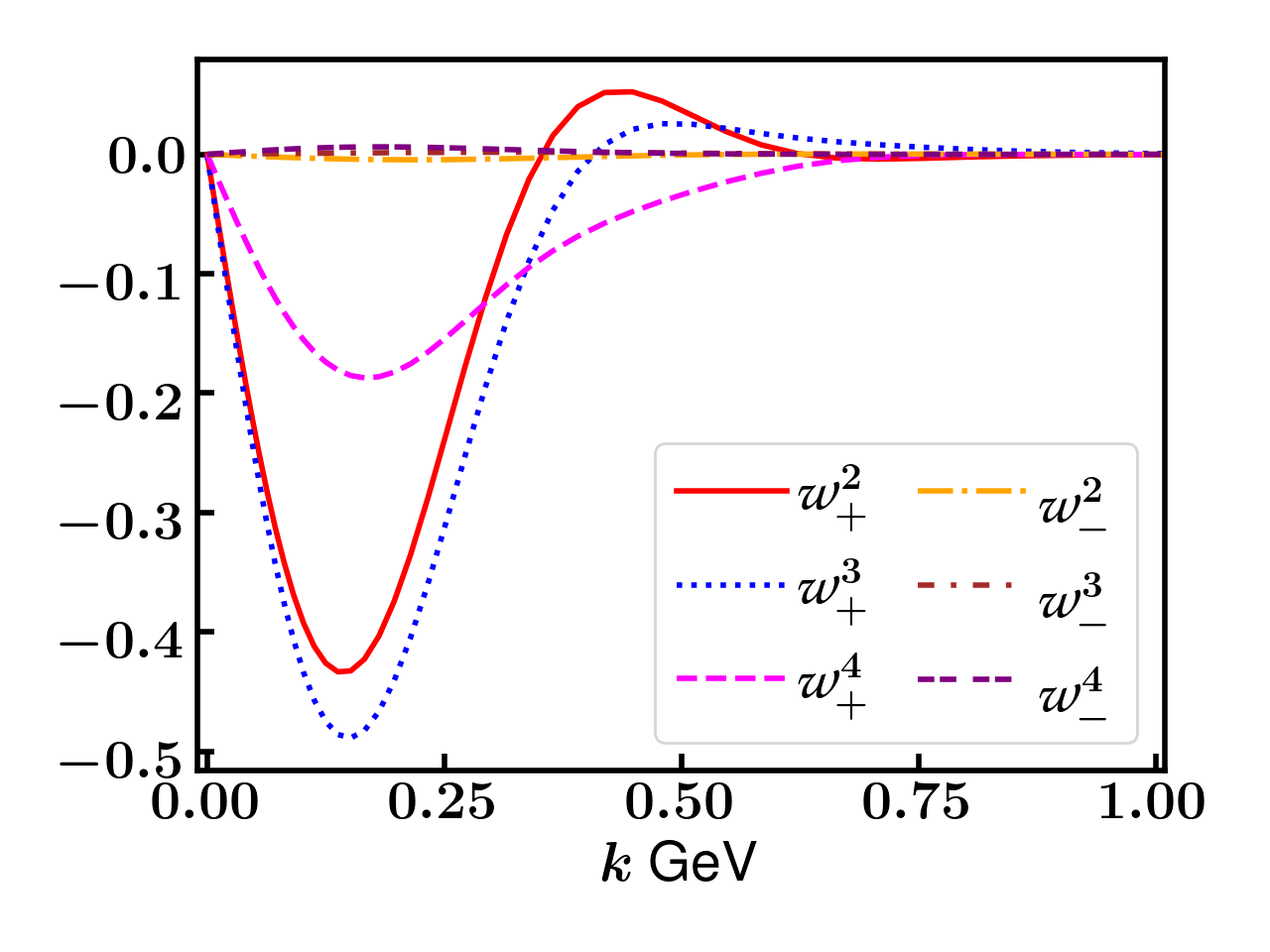} \\
\end{tabular}
\caption{
Zeroth Chebyshev moments -- Eq.\,\eqref{Chebyshev}.
\emph{Upper panels}.  Rest-frame $\mathsf S$-wave components in wave functions of the positive parity baryons: {\sf A} -- $\Delta(1232)\tfrac{3}{2}^+$; and {\sf B} -- $\Delta(1600)\tfrac{3}{2}^+$.
\emph{Lower panels}.  Rest-frame $\mathsf P$-wave components in wave functions of the negative parity baryons: {\sf C} -- $\Delta(1700)\tfrac{3}{2}^-$; and {\sf D} -- $\Delta(1940)\tfrac{3}{2}^-$.}
\end{figure*}

Today, numerical solutions of the quark gap equation can readily be obtained, but the utility of an algebraic form for $S(k)$ when calculations require the evaluation of numerous multidimensional integrals is clear.  An efficacious parametrisation has been used extensively:
{\allowdisplaybreaks
\begin{subequations}
\label{EqSSSV}
\begin{align}
\bar\sigma_S(x) & =  2\,\bar m \,{\cal F}(2 (x+\bar m^2)) \nonumber \\
& \quad + {\cal F}(b_1 x) \,{\cal F}(b_3 x) \,
\left[b_0 + b_2 {\cal F}(\epsilon x)\right]\,,\label{ssm} \\
\label{svm} \bar\sigma_V(x) & =  \frac{1}{x+\bar m^2}\, \left[ 1 - {\cal F}(2
(x+\bar m^2))\right]\,,
\end{align}
\end{subequations}}
\hspace*{-0.5\parindent}with $x=p^2/\lambda^2$, $\bar m$ = $m/\lambda$,
\begin{equation}
\label{defcalF}
{\cal F}(x)= \frac{1-\mbox{\rm e}^{-x}}{x}  \,,
\end{equation}
$\bar\sigma_S(x) = \lambda\,\sigma_S(k^2)$ and $\bar\sigma_V(x) =
\lambda^2\,\sigma_V(k^2)$.
The mass-scale, $\lambda=0.566\,$GeV, and
parameter values
\begin{equation}
\label{tableA}
\begin{array}{ccccc}
   \bar m& b_0 & b_1 & b_2 & b_3 \\\hline
   0.00897 & 0.131 & 2.90 & 0.603 & 0.185
\end{array}\;,
\end{equation}
associated with Eqs.\,\eqref{EqSSSV} were fixed in a least-squares fit to light-meson observables \cite{Burden:1995ve, Hecht:2000xa}.  ($\epsilon=10^{-4}$ in Eq.\,(\ref{ssm}) acts only to decouple the large- and intermediate-$k^2$ domains.)

The dimensionless light-quark current-mass in Eq.\,(\ref{tableA}) corresponds to
$m=5.08\,{\rm MeV}$ 
and the parametrisation yields the following Euclidean constituent-quark mass, defined as the solution of $k^2=M^2(k^2)$:
$M^E = 0.33\,{\rm GeV}$.
The ratio $M^E/m = 65$ is one expression of DCSB in the parametrisation of $S(k)$.  It emphasises the marked enhancement of the dressed-quark mass function at infrared momenta.

The dressed-quark mass function generated by this parametrisation compares well with that computed using sophisticated gap equation kernels \cite[Fig.\,A.1]{Chen:2017pse}.

A propagator is associated with each quark+quark correlation in Fig.\,\ref{FigFaddeev}; and we use 
\begin{align}
D^{1^\pm}_{\mu\nu}(K) & = \left[ \delta_{\mu\nu} + \frac{K_\mu K_\nu}{m_{1^\pm}^2} \right]
 \frac{1}{m_{1^\pm}^2} \, {\mathpzc F}(k^2/\omega_{1^\pm}^2)\,.
\label{Eqqqprop}
\end{align}

Our propagator representations ensure that the quarks and diquarks are confined within the baryons, as appropriate for coloured objects: whilst the propagators are free-particle-like at spacelike momenta, they are pole-free on the timelike axis; and this is sufficient to ensure confinement via the violation of reflection positivity (see, \emph{e.g}., Ref.\,\cite[Sec.\,3]{Horn:2016rip}).


\subsection{Chebyshev moments of rest-frame wave functions}
\label{WaveFunction}
Illustrating the remarks in Sec.\,\ref{SecQuarkCore}, Figs.\,\hyperref[NewFigureDeltaPositiveS]{A.1.A--D} display the zeroth Chebyshev moments of selected components in $(\tfrac{3}{2},\tfrac{3}{2}^\pm)$-baryon rest-frame wave functions.  When zeros appear, they are typically located within the domain $ \tfrac{1}{3}{\rm fm}\lesssim \frac{1}{k} \lesssim \tfrac{1}{2}{\rm fm}$, \emph{i.e}., at length-scales smaller than the bound-state radii.

\begin{table}[thb]
\begin{AAtable}
\label{NewTableSPDN940}
\addtocounter{Atable}{1}
\end{AAtable}
\caption{
Proton -- canonical normalisation contributions broken into rest-frame quark+diquark orbital angular momentum components, defined with reference to the scheme described in Refs.\,\cite{Oettel:1998bk, Cloet:2007pi}.  Where there are numerical differences with Ref.\,\cite{Cloet:2007pi}, we consider the results here to be more reliable.}
\begin{center}
\begin{tabular*}
{\hsize}
{
c@{\extracolsep{0ptplus1fil}}
|c@{\extracolsep{0ptplus1fil}}
c@{\extracolsep{0ptplus1fil}}
c@{\extracolsep{0ptplus1fil}}
|c@{\extracolsep{0ptplus1fil}}
c@{\extracolsep{0ptplus1fil}}
c@{\extracolsep{0ptplus1fil}}
c@{\extracolsep{0ptplus1fil}}
|c@{\extracolsep{0ptplus1fil}}}\hline
			& $ \mathcal{S}_1 $ & $ \mathcal{A}_2 $ & $ \mathcal{B}_1 $ & $\mathcal{S}_2$ & $ \mathcal{A}_1 $ & $ \mathcal{B}_2 $ & $ \mathcal{C}_2 $ & $ \mathcal{C}_1 $ \\
			\hline
			$ \mathcal{S}_1\ $& \;0.39 & -0.01 & \;0.08\; & 0.05 & \;0.00 & \;0.00 & \;0.00\; & \;0.00 \\
			$ \mathcal{A}_2\ $& -0.01 & \;0.00 & -0.07\; & 0.00 & \;0.03 & -0.02 & \;0.04\; & -0.02 \\
			$ \mathcal{B}_1\ $& \;0.08 & -0.07 & -0.04\; & 0.00 & \;0.01 & \;0.00 & \;0.33\; & \;0.01 \\
			\hline
			$\mathcal{S}_2$\ & \;0.05 & \;0.00 & \;0.00\; & 0.12 & \;0.00 & \;0.00 & \;0.00\; & \;0.00 \\
			$ \mathcal{A}_1\ $& \;0.00 & \;0.03 & \;0.01\; & 0.00 & \;0.00 & \;0.02 & -0.08\; & \;0.00 \\
			$ \mathcal{B}_2\ $& \;0.00 & -0.02 & \;0.00\; & 0.00 & \;0.02 & -0.01 & \;0.02\; & -0.01 \\
			$ \mathcal{C}_2\ $& \;0.00 & \;0.04 & \;0.33\; & 0.00 & -0.08 & \;0.02 & -0.24\; & \;0.04 \\
			\hline
			$ \mathcal{C}_1\ $& \;0.00 & -0.02 & \;0.01\; & 0.00 & \;0.00 & -0.01 & \;0.04\; & -0.01 \\\hline
\end{tabular*}
\end{center}
\end{table}

\begin{table}[thb]
\begin{AAtable}
\label{NewTableSPDD1232}
\addtocounter{Atable}{1}
\end{AAtable}
\caption{
$ \Delta(1232)\tfrac{3}{2}^+$ -- canonical normalisation contributions broken into rest-frame quark+$(1,1^+)$-diquark orbital angular momentum components, defined with reference to Table~\ref{TabL} and the basis in Eq.\,\eqref{AngularMomentum}.
 }
\begin{center}
\begin{tabular*}
{\hsize}
{
c@{\extracolsep{0ptplus1fil}}
|c@{\extracolsep{0ptplus1fil}}
|c@{\extracolsep{0ptplus1fil}}
c@{\extracolsep{0ptplus1fil}}
c@{\extracolsep{0ptplus1fil}}
|c@{\extracolsep{0ptplus1fil}}
c@{\extracolsep{0ptplus1fil}}
c@{\extracolsep{0ptplus1fil}}
|c@{\extracolsep{0ptplus1fil}}}\hline
			& $ \mathpzc{V}_1\ $ & $ \mathpzc{V}_2 $ & $ \mathpzc{V}_3 $ & $ \mathpzc{V}_4 $ & $ \mathpzc{V}_5 $ & $ \mathpzc{V}_6 $ & $ \mathpzc{V}_7 $ & $ \mathpzc{V}_8 $ \\
			\hline
			$ \mathpzc{V}_1\ $& \;3.90\; & -1.62 & -1.28 & -0.14$\ $ & -0.01 & 0.00 & 0.10$\ $ & 0.00 \\
			\hline
			$ \mathpzc{V}_2\ $& -1.69\; & \;1.77 & \;0.02 & \;0.04\; & -0.01 & \;0.00 & -0.03$\ $ & 0.00 \\
			$ \mathpzc{V}_3\ $& -1.27\; & \;0.03 & \;1.21 & \;0.06\; & \;0.02 & \;0.01 & -0.05$\ $ & 0.00 \\
			$ \mathpzc{V}_4\ $& -0.15\; & \;0.04 & \;0.06 & -0.06\; & \;0.01 & \;0.00 & \;0.04$\ $ & 0.00 \\
			\hline
			$ \mathpzc{V}_5\ $& -0.01\; & -0.01 & \;0.02 & \;0.01\; & -0.01 & \;0.00 & \;0.01$\ $ & 0.00 \\
			$ \mathpzc{V}_6\ $& \;0.00\; & \;0.00 & \;0.01 & \;0.00\; & \;0.00 & \;0.00 & -0.01$\ $ & 0.00 \\
			$ \mathpzc{V}_7\ $& \;0.10\; & -0.03 & -0.05 & \;0.04\; & \;0.01 & -0.01 & -0.08$\ $ & 0.00 \\
			\hline
			$ \mathpzc{V}_8\ $& \;0.00\; & \;0.00 & \;0.00 & \;0.00\; & \;0.00 & \;0.00 & \;0.00$\ $ & 0.00 \\\hline
\end{tabular*}
\end{center}
\end{table}


\subsection{Quark+diquark angular momentum}
\label{AAngular}
Using our solutions of the Faddeev equations for the Poincar\'e-covariant baryon wave functions, evaluated in the rest frame, we computed the contributions of various quark+diquark orbital angular momentum components to each baryon's canonical normalisation constant.  The results are recorded in 
Table~A.\hyperref[NewTableSPDN940]{1} for the nucleon and
Tables~A.\hyperref[NewTableSPDD1232]{2}\,--\,A.\hyperref[NewTableSPDD1940]{5}
for the $\Delta$-baryons.   It is from these tables that the images in Figs.\,\ref{FigNucleonL} and \ref{LFigures} are drawn.


\begin{table}[htb]
\begin{AAtable}
\label{NewTableSPDD1600}
\addtocounter{Atable}{1}
\end{AAtable}
\caption{
$ \Delta(1600)\tfrac{3}{2}^+$ -- canonical normalisation contributions broken into rest-frame quark+$(1,1^+)$-diquark orbital angular momentum components, defined with reference to Table~\ref{TabL} and the basis in Eq.\,\eqref{AngularMomentum}.
 }
\begin{center}
\begin{tabular*}
{\hsize}
{
c@{\extracolsep{0ptplus1fil}}
|c@{\extracolsep{0ptplus1fil}}
|c@{\extracolsep{0ptplus1fil}}
c@{\extracolsep{0ptplus1fil}}
c@{\extracolsep{0ptplus1fil}}
|c@{\extracolsep{0ptplus1fil}}
c@{\extracolsep{0ptplus1fil}}
c@{\extracolsep{0ptplus1fil}}
|c@{\extracolsep{0ptplus1fil}}}\hline
			& $ \mathpzc{V}_1\ $ & $ \mathpzc{V}_2 $ & $ \mathpzc{V}_3 $ & $ \mathpzc{V}_4 $ & $ \mathpzc{V}_5 $ & $ \mathpzc{V}_6 $ & $ \mathpzc{V}_7 $ & $ \mathpzc{V}_8 $ \\
			\hline
			$ \mathpzc{V}_1\ $& \;1.03$\ $ & \;0.00 & -0.05 & -0.02$\ $ & \;0.03 & -0.01 & -0.21$\ $ & \;0.07 \\
			\hline
			$ \mathpzc{V}_2\ $ & -0.01$\ $ & \;0.03 & -0.01 & -0.04$\ $ & \;0.08 & \;0.03 & \;0.04$\ $ & -0.02 \\
			$ \mathpzc{V}_3\ $& -0.07$\ $ & -0.01 & \;0.20 & -0.22$\ $ & -0.08 & \;0.04 & \;0.11$\ $ & -0.05 \\
			$ \mathpzc{V}_4\ $& -0.02$\ $ & -0.04 & -0.22 & -0.15$\ $ & -0.04 & \;0.02 & \;0.34$\ $ & -0.14 \\
			\hline
			$ \mathpzc{V}_5\ $& \;0.03$\ $ & \;0.08 & -0.09 & -0.04$\ $ & \;0.48 & \;0.00 & -0.02$\ $ & -0.03 \\
			$ \mathpzc{V}_6\ $& -0.01$\ $ & \;0.03 & \;0.05 & \;0.03$\ $ & \;0.00 & -0.02 & -0.12$\ $ & \;0.08 \\
			$ \mathpzc{V}_7\ $& -0.21$\ $ & \;0.04 & \;0.10 & \;0.33$\ $ & -0.02 & -0.11 & -0.30$\ $ & \;0.13 \\
			\hline
			$ \mathpzc{V}_8\ $& \;0.08$\ $ & -0.02 & -0.05 & -0.14$\ $ & -0.03 & \;0.08 & \;0.13$\ $ & -0.02 \\\hline
\end{tabular*}
\end{center}
\end{table}


\begin{table}[!h]
\begin{AAtable}
\label{NewTableSPDD1700}
\addtocounter{Atable}{1}
\end{AAtable}
\caption{
$ \Delta(1700)\tfrac{3}{2}^-$ -- canonical normalisation contributions broken into rest-frame quark+$(1,1^+)$-diquark orbital angular momentum components, defined with reference to Table~\ref{TabL} and the basis in Eq.\,\eqref{AngularMomentum}.
 }
\begin{center}
\begin{tabular*}
{\hsize}
{
c@{\extracolsep{0ptplus1fil}}
|c@{\extracolsep{0ptplus1fil}}
|c@{\extracolsep{0ptplus1fil}}
c@{\extracolsep{0ptplus1fil}}
c@{\extracolsep{0ptplus1fil}}
|c@{\extracolsep{0ptplus1fil}}
c@{\extracolsep{0ptplus1fil}}
c@{\extracolsep{0ptplus1fil}}
|c@{\extracolsep{0ptplus1fil}}}\hline
			& $ \mathpzc{V}_1\ $ & $ \mathpzc{V}_2 $ & $ \mathpzc{V}_3 $ & $ \mathpzc{V}_4 $ & $ \mathpzc{V}_5 $ & $ \mathpzc{V}_6 $ & $ \mathpzc{V}_7 $ & $ \mathpzc{V}_8 $ \\
			\hline
			$ \mathpzc{V}_1\ $& \;0.03$\ $ & \;0.02 & -0.05 & -0.16$\ $ & \;0.03 & \;0.01 & \;0.08$\ $ & -0.04 \\
			\hline
			$ \mathpzc{V}_2\ $ & \;0.02$\ $ & \;0.04 & \;0.00 & \;0.02$\ $ & -0.02 & -0.03 & \;0.00$\ $ & \;0.00 \\
			$ \mathpzc{V}_3\ $& -0.05$\ $ & \;0.00 & -0.59 & \;0.03$\ $ & -0.10 & \;0.00 & -0.11$\ $ & \;0.03 \\
			$ \mathpzc{V}_4\ $& -0.16$\ $ & \;0.02 & \;0.03 & \;2.36$\ $ & -0.25 & -0.48 & -0.17$\ $ & \;0.22 \\
			\hline
			$ \mathpzc{V}_5\ $& \;0.03$\ $ & -0.02 & -0.09 & -0.25$\ $ & \;0.26 & \;0.01 & \;0.12$\ $ & -0.13 \\
			$ \mathpzc{V}_6\ $& \;0.01$\ $ & -0.03 & \;0.00 & -0.48$\ $ & \;0.01 & \;1.04 & \;0.05$\ $ & -0.13 \\
			$ \mathpzc{V}_7\ $& \;0.08$\ $ & \;0.00 & -0.12 & -0.17$\ $ & \;0.12 & \;0.05 & -0.01$\ $ & -0.11 \\
			\hline
			$ \mathpzc{V}_8\ $& -0.04$\ $ & \;0.00 & \;0.03 & \;0.22$\ $ & -0.13 & -0.13 & -0.11$\ $ & \;0.20 \\\hline
\end{tabular*}
\end{center}
\end{table}

\begin{table}[!h]
\begin{AAtable}
\label{NewTableSPDD1940}
\addtocounter{Atable}{1}
\end{AAtable}
\caption{\label{TabSPDD1940}
$ \Delta(1940)\tfrac{3}{2}^-$ -- canonical normalisation contributions broken into rest-frame quark+$(1,1^+)$-diquark orbital angular momentum components, defined with reference to Table~\ref{TabL} and the basis in Eq.\,\eqref{AngularMomentum}.
 }
\begin{center}
\begin{tabular*}
{\hsize}
{
c@{\extracolsep{0ptplus1fil}}
|c@{\extracolsep{0ptplus1fil}}
|c@{\extracolsep{0ptplus1fil}}
c@{\extracolsep{0ptplus1fil}}
c@{\extracolsep{0ptplus1fil}}
|c@{\extracolsep{0ptplus1fil}}
c@{\extracolsep{0ptplus1fil}}
c@{\extracolsep{0ptplus1fil}}
|c@{\extracolsep{0ptplus1fil}}}\hline
			& $ \mathpzc{V}_1\ $ & $ \mathpzc{V}_2 $ & $ \mathpzc{V}_3 $ & $ \mathpzc{V}_4 $ & $ \mathpzc{V}_5 $ & $ \mathpzc{V}_6 $ & $ \mathpzc{V}_7 $ & $ \mathpzc{V}_8 $ \\
			\hline
			$ \mathpzc{V}_1\ $& \;0.77\; & -0.03 & -0.03 & -0.12$\ $ & \;0.00 & \;0.01 & \;0.01$\ $ & -0.01 \\
			\hline
			$ \mathpzc{V}_2\ $ & -0.03\; & \;0.27 & \;0.00 & \;0.00$\ $ & \;0.00 & \;0.00 & \;0.00$\ $ & \;0.00 \\
			$ \mathpzc{V}_3\ $& -0.03\; & \;0.00 & \;0.15 & \;0.02$\ $ & -0.12 & -0.01 & -0.04$\ $ & \;0.02 \\
			$ \mathpzc{V}_4\ $& -0.12\; & \;0.00 & \;0.01 & \;0.26$\ $ & \;0.02 & -0.02 & \;0.01$\ $ & \;0.01 \\
			\hline
			$ \mathpzc{V}_5\ $& \;0.00\; & \;0.00 & -0.12 & \;0.02$\ $ & \;0.11 & \;0.00 & \;0.03$\ $ & -0.04 \\
			$ \mathpzc{V}_6\ $& \;0.01\; & \;0.00 & -0.01 & -0.02$\ $ & \;0.00 & \;0.02 & -0.01$\ $ & \;0.02 \\
			$ \mathpzc{V}_7\ $& \;0.01\; & \;0.00 & -0.04 & \;0.01$\ $ & \;0.03 & -0.02 & -0.01$\ $ & -0.02 \\
			\hline
			$ \mathpzc{V}_8\ $& -0.01\; & \;0.00 & \;0.02 & \;0.01$\ $ & -0.04 & \;0.02 & -0.02$\ $ & \;0.04 \\\hline
\end{tabular*}
\end{center}
\end{table}

\newpage

$\,$

\newpage


\begin{thebibliography}{82}
\providecommand{\natexlab}[1]{#1}
\providecommand{\url}[1]{\texttt{#1}}
\providecommand{\urlprefix}{URL }
\expandafter\ifx\csname urlstyle\endcsname\relax
  \providecommand{\doi}[1]{doi:\discretionary{}{}{}#1}\else
  \providecommand{\doi}[1]{doi:\discretionary{}{}{}\begingroup
  \urlstyle{rm}\url{#1}\endgroup}\fi
\providecommand{\bibinfo}[2]{#2}

\bibitem[{Gell-Mann(1972)}]{Gell-Mann:2015noa}
\bibinfo{author}{M.~Gell-Mann}, \bibinfo{title}{{Quarks}},
  \bibinfo{journal}{Acta Phys. Austriaca Suppl.} \bibinfo{volume}{9}
  (\bibinfo{year}{1972}) \bibinfo{pages}{733--761}.

\bibitem[{Capstick and Roberts(2000)}]{Capstick:2000qj}
\bibinfo{author}{S.~Capstick}, \bibinfo{author}{W.~Roberts},
  \bibinfo{title}{{Quark models of baryon masses and decays}},
  \bibinfo{journal}{Prog. Part. Nucl. Phys.} \bibinfo{volume}{45}
  (\bibinfo{year}{2000}) \bibinfo{pages}{S241--S331}.

\bibitem[{Crede and Roberts(2013)}]{Crede:2013kia}
\bibinfo{author}{V.~Crede}, \bibinfo{author}{W.~Roberts},
  \bibinfo{title}{{Progress towards understanding baryon resonances}},
  \bibinfo{journal}{Rept. Prog. Phys.} \bibinfo{volume}{76}
  (\bibinfo{year}{2013}) \bibinfo{pages}{076301}.

\bibitem[{Giannini and Santopinto(2015)}]{Giannini:2015zia}
\bibinfo{author}{M.~M. Giannini}, \bibinfo{author}{E.~Santopinto},
  \bibinfo{title}{{The hypercentral Constituent Quark Model and its application
  to baryon properties}}, \bibinfo{journal}{Chin. J. Phys.}
  \bibinfo{volume}{53} (\bibinfo{year}{2015}) \bibinfo{pages}{020301}.

\bibitem[{Plessas(2015)}]{Plessas:2015mpa}
\bibinfo{author}{W.~Plessas}, \bibinfo{title}{{The constituent-quark model
  \textemdash{} Nowadays}}, \bibinfo{journal}{Int. J. Mod. Phys. A}
  \bibinfo{volume}{30}~(\bibinfo{number}{02}) (\bibinfo{year}{2015})
  \bibinfo{pages}{1530013}.

\bibitem[{Eichmann(2022)}]{Eichmann:2022zxn}
\bibinfo{author}{G.~Eichmann}, \bibinfo{title}{{Theory introduction to baryon
  spectroscopy -- arXiv:2202.13378 [hep-ph]}}, in: \bibinfo{booktitle}{{Baryons
  2021 -- 15th International Conference on the Structure of Baryons}},
  \bibinfo{pages}{Few Body Syst. \emph{in press}}, \bibinfo{year}{2022}.

\bibitem[{Zyla et~al.(2020)}]{Zyla:2020zbs}
\bibinfo{author}{P.~Zyla}, et~al., \bibinfo{title}{{Review of Particle
  Physics}}, \bibinfo{journal}{PTEP} \bibinfo{volume}{2020}
  (\bibinfo{year}{2020}) \bibinfo{pages}{083C01}.

\bibitem[{Burkert and Roberts(2019)}]{Burkert:2019bhp}
\bibinfo{author}{V.~D. Burkert}, \bibinfo{author}{C.~D. Roberts},
  \bibinfo{title}{{Colloquium: Roper resonance: Toward a solution to the
  fifty-year puzzle}}, \bibinfo{journal}{Rev. Mod. Phys.} \bibinfo{volume}{91}
  (\bibinfo{year}{2019}) \bibinfo{pages}{011003}.

\bibitem[{Brodsky et~al.(2022)Brodsky, Deur, and Roberts}]{Brodsky:2022fqy}
\bibinfo{author}{S.~J. Brodsky}, \bibinfo{author}{A.~Deur},
  \bibinfo{author}{C.~D. Roberts}, \bibinfo{title}{{Artificial Dynamical
  Effects in Quantum Field Theory -- arXiv:2202.06051 [hep-ph]}},
  \bibinfo{journal}{Nature Reviews Physics}  (\bibinfo{year}{2022})
  \bibinfo{pages}{\emph{in press}}.

\bibitem[{Holt and Roberts(2010)}]{Holt:2010vj}
\bibinfo{author}{R.~J. Holt}, \bibinfo{author}{C.~D. Roberts},
  \bibinfo{title}{{Distribution Functions of the Nucleon and Pion in the
  Valence Region}}, \bibinfo{journal}{Rev. Mod. Phys.} \bibinfo{volume}{82}
  (\bibinfo{year}{2010}) \bibinfo{pages}{2991--3044}.

\bibitem[{Carman et~al.(2020)Carman, Joo, and Mokeev}]{Carman:2020qmb}
\bibinfo{author}{D.~Carman}, \bibinfo{author}{K.~Joo},
  \bibinfo{author}{V.~Mokeev}, \bibinfo{title}{{Strong QCD Insights from
  Excited Nucleon Structure Studies with CLAS and CLAS12}},
  \bibinfo{journal}{Few Body Syst.} \bibinfo{volume}{61} (\bibinfo{year}{2020})
  \bibinfo{pages}{29}.

\bibitem[{Brodsky et~al.(2020)}]{Brodsky:2020vco}
\bibinfo{author}{S.~J. Brodsky}, et~al., \bibinfo{title}{{Strong QCD from
  Hadron Structure Experiments}}, \bibinfo{journal}{Intern. J. Mod. Phys. E}
  \bibinfo{volume}{124} (\bibinfo{year}{2020}) \bibinfo{pages}{2030006}.

\bibitem[{Mokeev and Carman(2022)}]{Mokeev:2022xfo}
\bibinfo{author}{V.~I. Mokeev}, \bibinfo{author}{D.~S. Carman},
  \bibinfo{title}{{Photo- and Electrocouplings of Nucleon Resonances --
  arXiv:2202.04180 [nucl-ex]}}, in: \bibinfo{booktitle}{{Baryons 2021 -- 15th
  International Conference on the Structure of Baryons}}, \bibinfo{pages}{Few
  Body Syst. \emph{in press}}, \bibinfo{year}{2022}.

\bibitem[{Coester(1992)}]{Coester:1992cg}
\bibinfo{author}{F.~Coester}, \bibinfo{title}{{Null plane dynamics of particles
  and fields}}, \bibinfo{journal}{Prog. Part. Nucl. Phys.} \bibinfo{volume}{29}
  (\bibinfo{year}{1992}) \bibinfo{pages}{1--32}.

\bibitem[{Edwards et~al.(2011)Edwards, Dudek, Richards, and
  Wallace}]{Edwards:2011jj}
\bibinfo{author}{R.~G. Edwards}, \bibinfo{author}{J.~J. Dudek},
  \bibinfo{author}{D.~G. Richards}, \bibinfo{author}{S.~J. Wallace},
  \bibinfo{title}{{Excited state baryon spectroscopy from lattice QCD}},
  \bibinfo{journal}{Phys. Rev. D} \bibinfo{volume}{84} (\bibinfo{year}{2011})
  \bibinfo{pages}{074508}.

\bibitem[{Fodor and Hoelbling(2012)}]{Fodor:2012gf}
\bibinfo{author}{Z.~Fodor}, \bibinfo{author}{C.~Hoelbling},
  \bibinfo{title}{{Light Hadron Masses from Lattice QCD}},
  \bibinfo{journal}{Rev. Mod. Phys.} \bibinfo{volume}{84}
  (\bibinfo{year}{2012}) \bibinfo{pages}{449}.

\bibitem[{Eichmann et~al.(2016{\natexlab{a}})Eichmann, Sanchis-Alepuz,
  Williams, Alkofer, and Fischer}]{Eichmann:2016yit}
\bibinfo{author}{G.~Eichmann}, \bibinfo{author}{H.~Sanchis-Alepuz},
  \bibinfo{author}{R.~Williams}, \bibinfo{author}{R.~Alkofer},
  \bibinfo{author}{C.~S. Fischer}, \bibinfo{title}{{Baryons as relativistic
  three-quark bound states}}, \bibinfo{journal}{Prog. Part. Nucl. Phys.}
  \bibinfo{volume}{91} (\bibinfo{year}{2016}{\natexlab{a}})
  \bibinfo{pages}{1--100}.

\bibitem[{Qin and Roberts(2020)}]{Qin:2020rad}
\bibinfo{author}{S.-X. Qin}, \bibinfo{author}{C.~D. Roberts},
  \bibinfo{title}{{Impressions of the Continuum Bound State Problem in QCD}},
  \bibinfo{journal}{Chin. Phys. Lett.}
  \bibinfo{volume}{37}~(\bibinfo{number}{12}) (\bibinfo{year}{2020})
  \bibinfo{pages}{121201}.

\bibitem[{Eichmann et~al.(2010)Eichmann, Alkofer, Krassnigg, and
  Nicmorus}]{Eichmann:2009qa}
\bibinfo{author}{G.~Eichmann}, \bibinfo{author}{R.~Alkofer},
  \bibinfo{author}{A.~Krassnigg}, \bibinfo{author}{D.~Nicmorus},
  \bibinfo{title}{{Nucleon mass from a covariant three-quark Faddeev
  equation}}, \bibinfo{journal}{Phys. Rev. Lett.} \bibinfo{volume}{104}
  (\bibinfo{year}{2010}) \bibinfo{pages}{201601}.

\bibitem[{Eichmann(2011)}]{Eichmann:2011vu}
\bibinfo{author}{G.~Eichmann}, \bibinfo{title}{{Nucleon electromagnetic form
  factors from the covariant Faddeev equation}}, \bibinfo{journal}{Phys. Rev.
  D} \bibinfo{volume}{84} (\bibinfo{year}{2011}) \bibinfo{pages}{014014}.

\bibitem[{Wang et~al.(2018)Wang, Qin, Roberts, and Schmidt}]{Wang:2018kto}
\bibinfo{author}{Q.-W. Wang}, \bibinfo{author}{S.-X. Qin},
  \bibinfo{author}{C.~D. Roberts}, \bibinfo{author}{S.~M. Schmidt},
  \bibinfo{title}{{Proton tensor charges from a Poincar{\'e}-covariant Faddeev
  equation}}, \bibinfo{journal}{Phys. Rev. D} \bibinfo{volume}{98}
  (\bibinfo{year}{2018}) \bibinfo{pages}{054019}.

\bibitem[{Qin et~al.(2019)Qin, Roberts, and Schmidt}]{Qin:2019hgk}
\bibinfo{author}{S.-X. Qin}, \bibinfo{author}{C.~D. Roberts},
  \bibinfo{author}{S.~M. Schmidt}, \bibinfo{title}{{Spectrum of light- and
  heavy-baryons}}, \bibinfo{journal}{Few Body Syst.} \bibinfo{volume}{60}
  (\bibinfo{year}{2019}) \bibinfo{pages}{26}.

\bibitem[{Qin et~al.(2013)Qin, Chang, Liu, Roberts, and Schmidt}]{Qin:2013mta}
\bibinfo{author}{S.-X. Qin}, \bibinfo{author}{L.~Chang}, \bibinfo{author}{Y.-X.
  Liu}, \bibinfo{author}{C.~D. Roberts}, \bibinfo{author}{S.~M. Schmidt},
  \bibinfo{title}{{Practical corollaries of transverse Ward-Green-Takahashi
  identities}}, \bibinfo{journal}{Phys.\ Lett.\ B} \bibinfo{volume}{722}
  (\bibinfo{year}{2013}) \bibinfo{pages}{384--388}, ISSN
  \bibinfo{issn}{0370-2693}.

\bibitem[{Qin et~al.(2014)Qin, Roberts, and Schmidt}]{Qin:2014vya}
\bibinfo{author}{S.-X. Qin}, \bibinfo{author}{C.~D. Roberts},
  \bibinfo{author}{S.~M. Schmidt}, \bibinfo{title}{{Ward-Green-Takahashi
  identities and the axial-vector vertex}}, \bibinfo{journal}{Phys. Lett. B}
  \bibinfo{volume}{733} (\bibinfo{year}{2014}) \bibinfo{pages}{202--208}.

\bibitem[{Binosi et~al.(2016)Binosi, Chang, Qin, Papavassiliou, and
  Roberts}]{Binosi:2016rxz}
\bibinfo{author}{D.~Binosi}, \bibinfo{author}{L.~Chang}, \bibinfo{author}{S.-X.
  Qin}, \bibinfo{author}{J.~Papavassiliou}, \bibinfo{author}{C.~D. Roberts},
  \bibinfo{title}{{Symmetry preserving truncations of the gap and
  Bethe-Salpeter equations}}, \bibinfo{journal}{Phys. Rev. D}
  \bibinfo{volume}{93} (\bibinfo{year}{2016}) \bibinfo{pages}{096010}.

\bibitem[{Qin and Roberts(2021)}]{Qin:2020jig}
\bibinfo{author}{S.-X. Qin}, \bibinfo{author}{C.~D. Roberts},
  \bibinfo{title}{{Resolving the Bethe-Salpeter kernel}},
  \bibinfo{journal}{Chin. Phys. Lett. \emph{Express}}
  \bibinfo{volume}{38}~(\bibinfo{number}{7}) (\bibinfo{year}{2021})
  \bibinfo{pages}{071201}.

\bibitem[{Cahill et~al.(1989)Cahill, Roberts, and Praschifka}]{Cahill:1988dx}
\bibinfo{author}{R.~T. Cahill}, \bibinfo{author}{C.~D. Roberts},
  \bibinfo{author}{J.~Praschifka}, \bibinfo{title}{{Baryon structure and QCD}},
  \bibinfo{journal}{Austral. J. Phys.} \bibinfo{volume}{42}
  (\bibinfo{year}{1989}) \bibinfo{pages}{129--145}.

\bibitem[{Burden et~al.(1989)Burden, Cahill, and Praschifka}]{Burden:1988dt}
\bibinfo{author}{C.~J. Burden}, \bibinfo{author}{R.~T. Cahill},
  \bibinfo{author}{J.~Praschifka}, \bibinfo{title}{{Baryon Structure and {QCD}:
  Nucleon Calculations}}, \bibinfo{journal}{Austral. J. Phys.}
  \bibinfo{volume}{42} (\bibinfo{year}{1989}) \bibinfo{pages}{147--159}.

\bibitem[{Reinhardt(1990)}]{Reinhardt:1989rw}
\bibinfo{author}{H.~Reinhardt}, \bibinfo{title}{{Hadronization of Quark Flavor
  Dynamics}}, \bibinfo{journal}{Phys. Lett. B} \bibinfo{volume}{244}
  (\bibinfo{year}{1990}) \bibinfo{pages}{316--326}.

\bibitem[{Efimov et~al.(1990)Efimov, Ivanov, and Lyubovitskij}]{Efimov:1990uz}
\bibinfo{author}{G.~V. Efimov}, \bibinfo{author}{M.~A. Ivanov},
  \bibinfo{author}{V.~E. Lyubovitskij}, \bibinfo{title}{{Quark - diquark
  approximation of the three quark structure of baryons in the quark
  confinement model}}, \bibinfo{journal}{Z. Phys. C} \bibinfo{volume}{47}
  (\bibinfo{year}{1990}) \bibinfo{pages}{583--594}.

\bibitem[{Barabanov et~al.(2021)}]{Barabanov:2020jvn}
\bibinfo{author}{M.~Y. Barabanov}, et~al., \bibinfo{title}{{Diquark
  Correlations in Hadron Physics: Origin, Impact and Evidence}},
  \bibinfo{journal}{Prog. Part. Nucl. Phys.} \bibinfo{volume}{116}
  (\bibinfo{year}{2021}) \bibinfo{pages}{103835}.

\bibitem[{Anselmino et~al.(1993)Anselmino, Predazzi, Ekelin, Fredriksson, and
  Lichtenberg}]{Anselmino:1992vg}
\bibinfo{author}{M.~Anselmino}, \bibinfo{author}{E.~Predazzi},
  \bibinfo{author}{S.~Ekelin}, \bibinfo{author}{S.~Fredriksson},
  \bibinfo{author}{D.~B. Lichtenberg}, \bibinfo{title}{{Diquarks}},
  \bibinfo{journal}{Rev. Mod. Phys.} \bibinfo{volume}{65}
  (\bibinfo{year}{1993}) \bibinfo{pages}{1199--1234}.

\bibitem[{Aznauryan et~al.(2011)Aznauryan, Burkert, Lee, and
  Mokeev}]{Aznauryan:2011ub}
\bibinfo{author}{I.~Aznauryan}, \bibinfo{author}{V.~Burkert},
  \bibinfo{author}{T.-S. Lee}, \bibinfo{author}{V.~Mokeev},
  \bibinfo{title}{{Results from the N* program at JLab}}, \bibinfo{journal}{J.
  Phys. Conf. Ser.} \bibinfo{volume}{299} (\bibinfo{year}{2011})
  \bibinfo{pages}{012008}.

\bibitem[{Chen et~al.(2018)Chen, El-Bennich, Roberts, Schmidt, Segovia, and
  Wan}]{Chen:2017pse}
\bibinfo{author}{C.~Chen}, \bibinfo{author}{B.~El-Bennich},
  \bibinfo{author}{C.~D. Roberts}, \bibinfo{author}{S.~M. Schmidt},
  \bibinfo{author}{J.~Segovia}, \bibinfo{author}{S.~Wan},
  \bibinfo{title}{{Structure of the nucleon's low-lying excitations}},
  \bibinfo{journal}{Phys. Rev. D} \bibinfo{volume}{97} (\bibinfo{year}{2018})
  \bibinfo{pages}{034016}.

\bibitem[{Chang et~al.(2022)Chang, Gao, and Roberts}]{Chang:2022jri}
\bibinfo{author}{L.~Chang}, \bibinfo{author}{F.~Gao}, \bibinfo{author}{C.~D.
  Roberts}, \bibinfo{title}{{Parton distributions of light quarks and
  antiquarks in the proton -- arXiv:2201.07870 [hep-ph]}} .

\bibitem[{Lu et~al.(2022)Lu, Chang, Raya, Roberts, and
  Rodr\'\i{}guez-Quintero}]{Lu:2022cjx}
\bibinfo{author}{Y.~Lu}, \bibinfo{author}{L.~Chang}, \bibinfo{author}{K.~Raya},
  \bibinfo{author}{C.~D. Roberts},
  \bibinfo{author}{J.~Rodr\'\i{}guez-Quintero}, \bibinfo{title}{{Proton and
  pion distribution functions in counterpoint -- arXiv:2203.00753 [hep-ph]}} .

\bibitem[{Abrams et~al.(2022)}]{Abrams:2021xum}
\bibinfo{author}{D.~Abrams}, et~al., \bibinfo{title}{{Measurement of the
  Nucleon $F^n_2/F^p_2$ Structure Function Ratio by the Jefferson Lab MARATHON
  Tritium/Helium-3 Deep Inelastic Scattering Experiment -- arXiv:2104.05850
  [hep-ex]$\!$}}, \bibinfo{journal}{Phys. Rev. Lett.}  (\bibinfo{year}{2022})
  \bibinfo{pages}{\emph{in press}}.

\bibitem[{Cui et~al.(2022)Cui, Gao, Binosi, Chang, Roberts, and
  Schmidt}]{Cui:2021gzg}
\bibinfo{author}{Z.-F. Cui}, \bibinfo{author}{F.~Gao},
  \bibinfo{author}{D.~Binosi}, \bibinfo{author}{L.~Chang},
  \bibinfo{author}{C.~D. Roberts}, \bibinfo{author}{S.~M. Schmidt},
  \bibinfo{title}{{Valence quark ratio in the proton}}, \bibinfo{journal}{Chin.
  Phys. Lett. \emph{Express}} \bibinfo{volume}{39}~(\bibinfo{number}{04})
  (\bibinfo{year}{2022}) \bibinfo{pages}{041401}.

\bibitem[{Oettel et~al.(1998)Oettel, Hellstern, Alkofer, and
  Reinhardt}]{Oettel:1998bk}
\bibinfo{author}{M.~Oettel}, \bibinfo{author}{G.~Hellstern},
  \bibinfo{author}{R.~Alkofer}, \bibinfo{author}{H.~Reinhardt},
  \bibinfo{title}{{Octet and decuplet baryons in a covariant and confining
  diquark - quark model}}, \bibinfo{journal}{Phys. Rev. C} \bibinfo{volume}{58}
  (\bibinfo{year}{1998}) \bibinfo{pages}{2459--2477}.

\bibitem[{Cloet et~al.(l th)Cloet, Krassnigg, and Roberts}]{Cloet:2007pi}
\bibinfo{author}{I.~C. Cloet}, \bibinfo{author}{A.~Krassnigg},
  \bibinfo{author}{C.~D. Roberts}, \bibinfo{title}{{Dynamics, Symmetries and
  Hadron Properties}} \bibinfo{note}{In \emph{11th} \emph{International}
  \emph{Conference} \emph{on} {Meson-Nucleon Physics and} {\it the} {\it
  Structure} \emph{of} \emph{the} \emph{Nucleon} \emph{(MENU 2007)}, eds.\
  H.~Machner and S.~Krewald, paper 125}.

\bibitem[{Hilger et~al.(2017)Hilger, Gomez-Rocha, and
  Krassnigg}]{Hilger:2015ora}
\bibinfo{author}{T.~Hilger}, \bibinfo{author}{M.~Gomez-Rocha},
  \bibinfo{author}{A.~Krassnigg}, \bibinfo{title}{{Light-quarkonium spectra and
  orbital-angular-momentum decomposition in a
  Bethe\textendash{}Salpeter-equation approach}}, \bibinfo{journal}{Eur. Phys.
  J. C} \bibinfo{volume}{77}~(\bibinfo{number}{9}) (\bibinfo{year}{2017})
  \bibinfo{pages}{625}.

\bibitem[{Weinberg(1967)}]{Weinberg:1967kj}
\bibinfo{author}{S.~Weinberg}, \bibinfo{title}{{Precise relations between the
  spectra of vector and axial vector mesons}}, \bibinfo{journal}{Phys. Rev.
  Lett.} \bibinfo{volume}{18} (\bibinfo{year}{1967}) \bibinfo{pages}{507--509}.

\bibitem[{Chang and Roberts(2012)}]{Chang:2011ei}
\bibinfo{author}{L.~Chang}, \bibinfo{author}{C.~D. Roberts},
  \bibinfo{title}{{Tracing masses of ground-state light-quark mesons}},
  \bibinfo{journal}{Phys. Rev. C} \bibinfo{volume}{85} (\bibinfo{year}{2012})
  \bibinfo{pages}{052201(R)}.

\bibitem[{Williams et~al.(2016)Williams, Fischer, and
  Heupel}]{Williams:2015cvx}
\bibinfo{author}{R.~Williams}, \bibinfo{author}{C.~S. Fischer},
  \bibinfo{author}{W.~Heupel}, \bibinfo{title}{{Light mesons in QCD and
  unquenching effects from the 3PI effective action}}, \bibinfo{journal}{Phys.
  Rev. D} \bibinfo{volume}{93} (\bibinfo{year}{2016}) \bibinfo{pages}{034026}.

\bibitem[{Roberts(2020)}]{Roberts:2020hiw}
\bibinfo{author}{C.~D. Roberts}, \bibinfo{title}{{Empirical Consequences of
  Emergent Mass}}, \bibinfo{journal}{Symmetry} \bibinfo{volume}{12}
  (\bibinfo{year}{2020}) \bibinfo{pages}{1468}.

\bibitem[{Roberts et~al.(2021)Roberts, Richards, Horn, and
  Chang}]{Roberts:2021nhw}
\bibinfo{author}{C.~D. Roberts}, \bibinfo{author}{D.~G. Richards},
  \bibinfo{author}{T.~Horn}, \bibinfo{author}{L.~Chang},
  \bibinfo{title}{{Insights into the emergence of mass from studies of pion and
  kaon structure}}, \bibinfo{journal}{Prog. Part. Nucl. Phys.}
  \bibinfo{volume}{120} (\bibinfo{year}{2021}) \bibinfo{pages}{103883}.

\bibitem[{Aguilar et~al.(2022)Aguilar, Ferreira, and
  Papavassiliou}]{Aguilar:2021uwa}
\bibinfo{author}{A.~C. Aguilar}, \bibinfo{author}{M.~N. Ferreira},
  \bibinfo{author}{J.~Papavassiliou}, \bibinfo{title}{{Exploring smoking-gun
  signals of the Schwinger mechanism in QCD}}, \bibinfo{journal}{Phys. Rev. D}
  \bibinfo{volume}{105}~(\bibinfo{number}{1}) (\bibinfo{year}{2022})
  \bibinfo{pages}{014030}.

\bibitem[{Binosi(2022)}]{Binosi:2022djx}
\bibinfo{author}{D.~Binosi}, \bibinfo{title}{{Emergent Hadron Mass in Strong
  Dynamics -- arXiv:2203.00942 [hep-ph]}}, in: \bibinfo{booktitle}{{Baryons
  2021 -- 15th International Conference on the Structure of Baryons}},
  \bibinfo{pages}{Few Body Syst. \emph{in press}}, \bibinfo{year}{2022}.

\bibitem[{Horn and Roberts(2016)}]{Horn:2016rip}
\bibinfo{author}{T.~Horn}, \bibinfo{author}{C.~D. Roberts},
  \bibinfo{title}{{The pion: an enigma within the Standard Model}},
  \bibinfo{journal}{J. Phys. G.} \bibinfo{volume}{43} (\bibinfo{year}{2016})
  \bibinfo{pages}{073001}.

\bibitem[{Yin et~al.(2021)Yin, Cui, Roberts, and Segovia}]{Yin:2021uom}
\bibinfo{author}{P.-L. Yin}, \bibinfo{author}{Z.-F. Cui},
  \bibinfo{author}{C.~D. Roberts}, \bibinfo{author}{J.~Segovia},
  \bibinfo{title}{{Masses of positive- and negative-parity hadron
  ground-states, including those with heavy quarks}}, \bibinfo{journal}{Eur.
  Phys. J. C} \bibinfo{volume}{81}~(\bibinfo{number}{4}) (\bibinfo{year}{2021})
  \bibinfo{pages}{327}.

\bibitem[{Alkofer et~al.(2005)Alkofer, H{\"o}ll, Kloker, Krassnigg, and
  Roberts}]{Alkofer:2004yf}
\bibinfo{author}{R.~Alkofer}, \bibinfo{author}{A.~H{\"o}ll},
  \bibinfo{author}{M.~Kloker}, \bibinfo{author}{A.~Krassnigg},
  \bibinfo{author}{C.~D. Roberts}, \bibinfo{title}{{On nucleon electromagnetic
  form-factors}}, \bibinfo{journal}{Few Body Syst.} \bibinfo{volume}{37}
  (\bibinfo{year}{2005}) \bibinfo{pages}{1--31}.

\bibitem[{Roberts et~al.(2011)Roberts, Chang, Cloet, and
  Roberts}]{Roberts:2011cf}
\bibinfo{author}{H.~L.~L. Roberts}, \bibinfo{author}{L.~Chang},
  \bibinfo{author}{I.~C. Cloet}, \bibinfo{author}{C.~D. Roberts},
  \bibinfo{title}{{Masses of ground and excited-state hadrons}},
  \bibinfo{journal}{Few Body Syst.} \bibinfo{volume}{51} (\bibinfo{year}{2011})
  \bibinfo{pages}{1--25}.

\bibitem[{Segovia et~al.(2014)Segovia, Cloet, Roberts, and
  Schmidt}]{Segovia:2014aza}
\bibinfo{author}{J.~Segovia}, \bibinfo{author}{I.~C. Cloet},
  \bibinfo{author}{C.~D. Roberts}, \bibinfo{author}{S.~M. Schmidt},
  \bibinfo{title}{{Nucleon and $\Delta$ elastic and transition form factors}},
  \bibinfo{journal}{Few Body Syst.} \bibinfo{volume}{55} (\bibinfo{year}{2014})
  \bibinfo{pages}{1185--1222}.

\bibitem[{Guti\'errez-Guerrero et~al.(2021)Guti\'errez-Guerrero,
  Paredes-Torres, and Bashir}]{Gutierrez-Guerrero:2021rsx}
\bibinfo{author}{L.~X. Guti\'errez-Guerrero},
  \bibinfo{author}{G.~Paredes-Torres}, \bibinfo{author}{A.~Bashir},
  \bibinfo{title}{{Mesons and baryons: Parity partners}},
  \bibinfo{journal}{Phys. Rev. D} \bibinfo{volume}{104}~(\bibinfo{number}{9})
  (\bibinfo{year}{2021}) \bibinfo{pages}{094013}.

\bibitem[{Arp(1998)}]{Arpack}
\bibinfo{note}{R.~B.~Lehoucq, D.~C.~Sorensen and C.~Yang, {\em ARPACK Users'
  Guide: Solution of Large-Scale Eigenvalue Problems with Implicitly Restarted
  Arnoldi Methods\/} (Society for Industrial \& Applied Mathematics)},
  \bibinfo{year}{1998}.

\bibitem[{Qiu(2021)}]{SPECTRA}
\bibinfo{author}{Y.~Qiu}, \bibinfo{note}{{\em Sparse Eigenvalue Computation
  Toolkit as a Redesigned ARPACK (SPECTRA)\/}
  (https://spectralib.org/index.html)}, \bibinfo{year}{2021}.

\bibitem[{Julia-Diaz et~al.(2007)Julia-Diaz, Lee, Matsuyama, and
  Sato}]{JuliaDiaz:2007kz}
\bibinfo{author}{B.~Julia-Diaz}, \bibinfo{author}{T.~S.~H. Lee},
  \bibinfo{author}{A.~Matsuyama}, \bibinfo{author}{T.~Sato},
  \bibinfo{title}{{Dynamical coupled-channel model of pi N scattering in the $
  W \leq 2$-GeV nucleon resonance region}}, \bibinfo{journal}{Phys. Rev. C}
  \bibinfo{volume}{76} (\bibinfo{year}{2007}) \bibinfo{pages}{065201}.

\bibitem[{Suzuki et~al.(2010)Suzuki, Julia-Diaz, Kamano, Lee, Matsuyama, and
  Sato}]{Suzuki:2009nj}
\bibinfo{author}{N.~Suzuki}, \bibinfo{author}{B.~Julia-Diaz},
  \bibinfo{author}{H.~Kamano}, \bibinfo{author}{T.~S.~H. Lee},
  \bibinfo{author}{A.~Matsuyama}, \bibinfo{author}{T.~Sato},
  \bibinfo{title}{{Disentangling the Dynamical Origin of P-11 Nucleon
  Resonances}}, \bibinfo{journal}{Phys. Rev. Lett.} \bibinfo{volume}{104}
  (\bibinfo{year}{2010}) \bibinfo{pages}{042302}.

\bibitem[{R{\"o}nchen et~al.(2013)R{\"o}nchen, D{\"o}ring, Huang, Haberzettl,
  Haidenbauer, Hanhart, Krewald, Meissner, and Nakayama}]{Ronchen:2012eg}
\bibinfo{author}{D.~R{\"o}nchen}, \bibinfo{author}{M.~D{\"o}ring},
  \bibinfo{author}{F.~Huang}, \bibinfo{author}{H.~Haberzettl},
  \bibinfo{author}{J.~Haidenbauer}, \bibinfo{author}{C.~Hanhart},
  \bibinfo{author}{S.~Krewald}, \bibinfo{author}{U.~G. Meissner},
  \bibinfo{author}{K.~Nakayama}, \bibinfo{title}{{Coupled-channel dynamics in
  the reactions $\pi N \to \pi N$, $\eta N$, $K\Lambda$, $K\Sigma$}},
  \bibinfo{journal}{Eur. Phys. J. A} \bibinfo{volume}{49}
  (\bibinfo{year}{2013}) \bibinfo{pages}{44}.

\bibitem[{Kamano et~al.(2013)Kamano, Nakamura, Lee, and Sato}]{Kamano:2013iva}
\bibinfo{author}{H.~Kamano}, \bibinfo{author}{S.~X. Nakamura},
  \bibinfo{author}{T.~S.~H. Lee}, \bibinfo{author}{T.~Sato},
  \bibinfo{title}{{Nucleon resonances within a dynamical coupled-channels model
  of $\pi N$ and $\gamma N$ reactions}}, \bibinfo{journal}{Phys. Rev. C}
  \bibinfo{volume}{88} (\bibinfo{year}{2013}) \bibinfo{pages}{035209}.

\bibitem[{Eichmann et~al.(2008)Eichmann, Alkofer, Cloet, Krassnigg, and
  Roberts}]{Eichmann:2008ae}
\bibinfo{author}{G.~Eichmann}, \bibinfo{author}{R.~Alkofer},
  \bibinfo{author}{I.~C. Cloet}, \bibinfo{author}{A.~Krassnigg},
  \bibinfo{author}{C.~D. Roberts}, \bibinfo{title}{{Perspective on
  rainbow-ladder truncation}}, \bibinfo{journal}{Phys. Rev. C}
  \bibinfo{volume}{77} (\bibinfo{year}{2008}) \bibinfo{pages}{042202(R)}.

\bibitem[{Eichmann et~al.(2009)Eichmann, Cloet, Alkofer, Krassnigg, and
  Roberts}]{Eichmann:2008ef}
\bibinfo{author}{G.~Eichmann}, \bibinfo{author}{I.~C. Cloet},
  \bibinfo{author}{R.~Alkofer}, \bibinfo{author}{A.~Krassnigg},
  \bibinfo{author}{C.~D. Roberts}, \bibinfo{title}{{Toward unifying the
  description of meson and baryon properties}}, \bibinfo{journal}{Phys. Rev. C}
  \bibinfo{volume}{79} (\bibinfo{year}{2009}) \bibinfo{pages}{012202(R)}.

\bibitem[{H{\"o}ll et~al.(2004)H{\"o}ll, Krassnigg, and Roberts}]{Holl:2004fr}
\bibinfo{author}{A.~H{\"o}ll}, \bibinfo{author}{A.~Krassnigg},
  \bibinfo{author}{C.~D. Roberts}, \bibinfo{title}{{Pseudoscalar meson radial
  excitations}}, \bibinfo{journal}{Phys. Rev. C} \bibinfo{volume}{70}
  (\bibinfo{year}{2004}) \bibinfo{pages}{042203(R)}.

\bibitem[{Li et~al.(2016)Li, Chang, Gao, Roberts, Schmidt, and
  Zong}]{Li:2016dzv}
\bibinfo{author}{B.-L. Li}, \bibinfo{author}{L.~Chang},
  \bibinfo{author}{F.~Gao}, \bibinfo{author}{C.~D. Roberts},
  \bibinfo{author}{S.~M. Schmidt}, \bibinfo{author}{H.-S. Zong},
  \bibinfo{title}{{Distribution amplitudes of radially-excited $\pi$ and $K$
  mesons}}, \bibinfo{journal}{Phys. Rev. D}
  \bibinfo{volume}{93}~(\bibinfo{number}{11}) (\bibinfo{year}{2016})
  \bibinfo{pages}{114033}.

\bibitem[{Qin et~al.(2012)Qin, Chang, Liu, Roberts, and Wilson}]{Qin:2011xq}
\bibinfo{author}{S.-X. Qin}, \bibinfo{author}{L.~Chang}, \bibinfo{author}{Y.-x.
  Liu}, \bibinfo{author}{C.~D. Roberts}, \bibinfo{author}{D.~J. Wilson},
  \bibinfo{title}{{Investigation of rainbow-ladder truncation for excited and
  exotic mesons}}, \bibinfo{journal}{Phys. Rev. C} \bibinfo{volume}{85}
  (\bibinfo{year}{2012}) \bibinfo{pages}{035202}.

\bibitem[{Eichmann and Nicmorus(2012)}]{Eichmann:2011aa}
\bibinfo{author}{G.~Eichmann}, \bibinfo{author}{D.~Nicmorus},
  \bibinfo{title}{{Nucleon to Delta electromagnetic transition in the
  Dyson-Schwinger approach}}, \bibinfo{journal}{Phys. Rev. D}
  \bibinfo{volume}{85} (\bibinfo{year}{2012}) \bibinfo{pages}{093004}.

\bibitem[{Lu et~al.(2019)Lu, Chen, Cui, Roberts, Schmidt, Segovia, and
  Zong}]{Lu:2019bjs}
\bibinfo{author}{Y.~Lu}, \bibinfo{author}{C.~Chen}, \bibinfo{author}{Z.-F.
  Cui}, \bibinfo{author}{C.~D. Roberts}, \bibinfo{author}{S.~M. Schmidt},
  \bibinfo{author}{J.~Segovia}, \bibinfo{author}{H.~S. Zong},
  \bibinfo{title}{{Transition form factors: $\gamma^\ast + p \to \Delta(1232)$,
  $\Delta(1600)$}}, \bibinfo{journal}{Phys. Rev. D} \bibinfo{volume}{100}
  (\bibinfo{year}{2019}) \bibinfo{pages}{034001}.

\bibitem[{Eichmann et~al.(2016{\natexlab{b}})Eichmann, Fischer, and
  Sanchis-Alepuz}]{Eichmann:2016hgl}
\bibinfo{author}{G.~Eichmann}, \bibinfo{author}{C.~S. Fischer},
  \bibinfo{author}{H.~Sanchis-Alepuz}, \bibinfo{title}{{Light baryons and their
  excitations}}, \bibinfo{journal}{Phys. Rev. D} \bibinfo{volume}{94}
  (\bibinfo{year}{2016}{\natexlab{b}}) \bibinfo{pages}{094033}.

\bibitem[{Qin et~al.(2018)Qin, Roberts, and Schmidt}]{Qin:2018dqp}
\bibinfo{author}{S.-X. Qin}, \bibinfo{author}{C.~D. Roberts},
  \bibinfo{author}{S.~M. Schmidt}, \bibinfo{title}{{Poincar{\'e}-covariant
  analysis of heavy-quark baryons}}, \bibinfo{journal}{Phys. Rev. D}
  \bibinfo{volume}{97} (\bibinfo{year}{2018}) \bibinfo{pages}{114017}.

\bibitem[{Mokeev(2022)}]{MokeevPrivate2022}
\bibinfo{author}{V.~I. Mokeev}, \bibinfo{title}{{private communication, 2022}}
  .

\bibitem[{Burkert et~al.(2003)Burkert, De~Vita, Battaglieri, Ripani, and
  Mokeev}]{Burkert:2002zz}
\bibinfo{author}{V.~D. Burkert}, \bibinfo{author}{R.~De~Vita},
  \bibinfo{author}{M.~Battaglieri}, \bibinfo{author}{M.~Ripani},
  \bibinfo{author}{V.~Mokeev}, \bibinfo{title}{{Single quark transition model
  analysis of electromagnetic nucleon resonance transitions in the [70,1-]
  supermultiplet}}, \bibinfo{journal}{Phys. Rev. C} \bibinfo{volume}{67}
  (\bibinfo{year}{2003}) \bibinfo{pages}{035204}.

\bibitem[{Dugger et~al.(2009)}]{CLAS:2009tyz}
\bibinfo{author}{M.~Dugger}, et~al., \bibinfo{title}{{$\pi^+$ photoproduction
  on the proton for photon energies from 0.725 to 2.875-GeV}},
  \bibinfo{journal}{Phys. Rev. C} \bibinfo{volume}{79} (\bibinfo{year}{2009})
  \bibinfo{pages}{065206}.

\bibitem[{Mokeev and Aznauryan(2014)}]{Mokeev:2013kka}
\bibinfo{author}{V.~I. Mokeev}, \bibinfo{author}{I.~G. Aznauryan},
  \bibinfo{title}{{Studies of $N^*$ structure from the CLAS meson
  electroproduction data}}, \bibinfo{journal}{Int. J. Mod. Phys. Conf. Ser.}
  \bibinfo{volume}{26} (\bibinfo{year}{2014}) \bibinfo{pages}{1460080}.

\bibitem[{Isupov et~al.(2017)}]{CLAS:2017fja}
\bibinfo{author}{E.~L. Isupov}, et~al., \bibinfo{title}{{Measurements of $e p
  \to e' \pi^+ \pi^- p'$ Cross Sections with CLAS at $1.40$ Gev $< W < 2.0$ GeV
  and $2.0$ GeV$^2$ $< Q^2 < 5.0$ GeV$^2$}}, \bibinfo{journal}{Phys. Rev. C}
  \bibinfo{volume}{96}~(\bibinfo{number}{2}) (\bibinfo{year}{2017})
  \bibinfo{pages}{025209}.

\bibitem[{Trivedi(2019)}]{Trivedi:2018rgo}
\bibinfo{author}{A.~Trivedi}, \bibinfo{title}{{Measurement of New Observables
  from the $\pi^+\pi^-$ p Electroproduction Off the Proton}},
  \bibinfo{journal}{Few Body Syst.} \bibinfo{volume}{60} (\bibinfo{year}{2019})
  \bibinfo{pages}{5}.

\bibitem[{Cui et~al.(2020)Cui, Chen, Binosi, de~Soto, Roberts,
  Rodr{\'{\i}}guez-Quintero, Schmidt, and Segovia}]{Cui:2020rmu}
\bibinfo{author}{Z.-F. Cui}, \bibinfo{author}{C.~Chen},
  \bibinfo{author}{D.~Binosi}, \bibinfo{author}{F.~de~Soto},
  \bibinfo{author}{C.~D. Roberts},
  \bibinfo{author}{J.~Rodr{\'{\i}}guez-Quintero}, \bibinfo{author}{S.~M.
  Schmidt}, \bibinfo{author}{J.~Segovia}, \bibinfo{title}{{Nucleon elastic form
  factors at accessible large spacelike momenta}}, \bibinfo{journal}{Phys. Rev.
  D} \bibinfo{volume}{102} (\bibinfo{year}{2020}) \bibinfo{pages}{014043}.

\bibitem[{Chen et~al.(2021{\natexlab{a}})Chen, Fischer, Roberts, and
  Segovia}]{Chen:2020wuq}
\bibinfo{author}{C.~Chen}, \bibinfo{author}{C.~S. Fischer},
  \bibinfo{author}{C.~D. Roberts}, \bibinfo{author}{J.~Segovia},
  \bibinfo{title}{{Form Factors of the Nucleon Axial Current}},
  \bibinfo{journal}{Phys. Lett. B} \bibinfo{volume}{815}
  (\bibinfo{year}{2021}{\natexlab{a}}) \bibinfo{pages}{136150}.

\bibitem[{Chen et~al.(2021{\natexlab{b}})Chen, Fischer, Roberts, and
  Segovia}]{Chen:2021guo}
\bibinfo{author}{C.~Chen}, \bibinfo{author}{C.~S. Fischer},
  \bibinfo{author}{C.~D. Roberts}, \bibinfo{author}{J.~Segovia},
  \bibinfo{title}{{Nucleon axial-vector and pseudoscalar form factors, and PCAC
  relations -- arXiv:2103.02054 [hep-ph]}} .

\bibitem[{Mosel(2016)}]{Mosel:2016cwa}
\bibinfo{author}{U.~Mosel}, \bibinfo{title}{{Neutrino Interactions with
  Nucleons and Nuclei: Importance for Long-Baseline Experiments}},
  \bibinfo{journal}{Ann. Rev. Nucl. Part. Sci.} \bibinfo{volume}{66}
  (\bibinfo{year}{2016}) \bibinfo{pages}{171--195}.

\bibitem[{Oliveira et~al.(2019)Oliveira, Silva, Skullerud, and
  Sternbeck}]{Oliveira:2018lln}
\bibinfo{author}{O.~Oliveira}, \bibinfo{author}{P.~J. Silva},
  \bibinfo{author}{J.-I. Skullerud}, \bibinfo{author}{A.~Sternbeck},
  \bibinfo{title}{{Quark propagator with two flavors of O(a)-improved Wilson
  fermions}}, \bibinfo{journal}{Phys. Rev. D} \bibinfo{volume}{99}
  (\bibinfo{year}{2019}) \bibinfo{pages}{094506}.

\bibitem[{Burden et~al.(1996)Burden, Roberts, and Thomson}]{Burden:1995ve}
\bibinfo{author}{C.~J. Burden}, \bibinfo{author}{C.~D. Roberts},
  \bibinfo{author}{M.~J. Thomson}, \bibinfo{title}{{Electromagnetic Form
  Factors of Charged and Neutral Kaons}}, \bibinfo{journal}{Phys. Lett. B}
  \bibinfo{volume}{371} (\bibinfo{year}{1996}) \bibinfo{pages}{163--168}.

\bibitem[{Hecht et~al.(2001)Hecht, Roberts, and Schmidt}]{Hecht:2000xa}
\bibinfo{author}{M.~B. Hecht}, \bibinfo{author}{C.~D. Roberts},
  \bibinfo{author}{S.~M. Schmidt}, \bibinfo{title}{{Valence-quark distributions
  in the pion}}, \bibinfo{journal}{Phys. Rev. C} \bibinfo{volume}{63}
  (\bibinfo{year}{2001}) \bibinfo{pages}{025213}.

\end{thebibliography}

\end{document}